\documentclass{article}

\usepackage{arxiv}
\usepackage[utf8]{inputenc} % allow utf-8 input
\usepackage[T1]{fontenc}    % use 8-bit T1 fonts
\usepackage{hyperref}       % hyperlinks
\usepackage{url}            % simple URL typesetting
\usepackage{booktabs}       % professional-quality tables
\usepackage{amsfonts}       % blackboard math symbols
\usepackage{nicefrac}       % compact symbols for 1/2, etc.
\usepackage{microtype}      % microtypography
\usepackage{cleveref}       % smart cross-referencing
\usepackage{graphicx}
\usepackage{doi}
\usepackage{lscape}
\usepackage{longtable}
\usepackage{placeins}
\usepackage{caption}
\usepackage{float}
\usepackage{wrapfig}
\usepackage[table,xcdraw]{xcolor}
\newcommand{\xrowht}[1]{\rule{0pt}{#1}}

\title{Introducing IHARDS-CNN: A Cutting-Edge Deep Learning Method for Human Activity Recognition Using Wearable Sensors}

\newbox{\orcid}\sbox{\orcid}{\includegraphics[scale=0.06 , keepaspectratio]{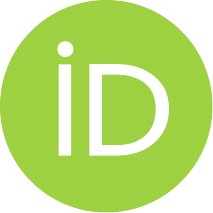}} 

\newif\ifuniqueAffiliation
\ifuniqueAffiliation % Standard variant of author block
...

\else
\usepackage{authblk}

\newbox{\orcid}\sbox{\orcid}{\includegraphics[scale=0.06 , keepaspectratio]{orcid.pdf}} 

\author[1,2]{\href{https://orcid.org/0009-0001-2006-7099}{\usebox{\orcid}\hspace{1mm}Nazanin Sedaghati}\thanks{\texttt{nsedaghati@iaut.ac.ir}}}
\author[1,2]{\href{https://orcid.org/0000-0002-6650-3538}{\usebox{\orcid}
\hspace{1mm}Masoud Kargar*}\thanks{\texttt{kargar@iaut.ac.ir}}}
\author[1,2]{\href{https://orcid.org/0009-0004-3274-8862}{\usebox{\orcid}\hspace{1mm}Sina Abbaskhani}\thanks{\texttt{sina.abbaskhani@iaut.ac.ir}}}

\affil[1]{Department of Computer Engineering, Islamic Azad University, Tabriz Branch, Iran}
\affil[2]{Robotics and Soft Technologies Research Center, Islamic Azad University, Tabriz Branch, Iran}

\fi

\renewcommand{\shorttitle}{\textit{arXiv} }

\hypersetup{
pdftitle={MULTISTAGE NON-DETERMINISTIC CLASSIFICATION USING
SECONDARY CONCEPT GRAPHS AND GRAPH
CONVOLUTIONAL NETWORKS FOR HIGH-LEVEL FEATURE
EXTRACTION},
pdfsubject={q-bio.NC, q-bio.QM},
pdfauthor={David S.~Hippocampus, Elias D.~Striatum},
pdfkeywords={First keyword, Second keyword, More},
}

\renewcommand{\shorttitle}{IHARDS-CNN, for Human Activity Recognition}

\begin{document}
\maketitle

\begin{abstract}
	Human activity recognition, facilitated by smart devices, has recently garnered significant attention. Deep learning algorithms have become pivotal in daily activities, sports, and healthcare. Nevertheless, addressing the challenge of extracting features from sensor data processing necessitates the utilization of diverse algorithms in isolation, subsequently transforming them into a standard mode. 
This research introduces a novel approach called IHARDS-CNN, amalgamating data from three distinct datasets (UCI-HAR, WISDM, and KU-HAR) for human activity recognition. The data collected from sensors embedded in smartwatches or smartphones encompass five daily activity classes. This study initially outlines the dataset integration approach, follows with a comprehensive statistical analysis, and assesses dataset accuracy. The proposed methodology employs a one-dimensional deep convolutional neural network for classification. Compared to extant activity recognition methods, this approach stands out for its high speed, reduced detection steps, and absence of the need to aggregate classified results. Despite fewer detection steps, empirical results demonstrate an impressive accuracy of nearly 100\%, marking it the highest among existing methods. Evaluation outcomes further highlight superior classification performance when compared to analogous architectures. 
\end{abstract}

% keywords can be removed

\section{Introduction}
In recent years, extensive research has focused on explaining the intricate relationship between health and biometric behavioral data of people\cite{impedovo2022ieee}. These researches primarily center on data collected via wearable sensors\cite{mekruksavanich2021biometric}. Treatment in medical science to diagnose chronic diseases and continuous monitoring of patients' daily activities has increased\cite{li2023human}.\\
Recent research endeavors have emphasized human action recognition and the diagnostic identification of human activities\cite{alghyaline2019real}. Human action recognition aims to gain a deeper understanding of behavior analysis, with machine vision as a key driver\cite{qing2023mar}. In contrast, the objective of human activity recognition involves recognizing and analyzing behavioral patterns across a broader spectrum of activities\cite{choudhury2023adaptive}.\\
Within human action recognition, two overarching categories exist: visual and non-visual. Notable examples of specialized datasets in the visual (RGB colors) type include UCF101, HMDB51, and KINECTIS-400 \cite{sun2022human},\cite{fu2016human} We can employ various methodologies for action recognition, including infrared, hyper-point, event flow, sound, radar acceleration, and encryption.\\
Human activity recognition based on data collection can be categorized into two main approaches: vision-based and sensor-based\cite{sarkar2023human}. However, vision-based data collection necessitates camera-equipped devices for capturing the surrounding environment, which can pose security concerns \cite{gholamiangonabadi2020deep}. On the other hand, sensor-based data collection relies on embedded sensors in wearable devices for Human Activity Recognition (HAR), employing accelerometers and gyroscopes \cite{mekruksavanich2021biometric}. Various sensors, such as environmental and wearable, are used for HAR detection.\\
From the perspective of activity complexity, HAR can be classified into three groups: short-duration, simple, and complex activities \cite{hassan2018robust}. Short-duration activities include transitioning from sitting to standing while walking is a simple activity. Complex activities emerge from combinations of interactions with people or objects.\\
Prominent datasets in human activity recognition encompass UCI-HAR, WISDM, PAMAP2, and MHEALTH \cite{ramanujam2021human}. The display of health and fitness data analysis and daily activity tracking is paramount in human activity recognition. Daily activities span stair climbing, walking, talking, eating, and lying down\cite{ismail2023auto}. Recognizing human activities finds applications in calorie calculation, eldercare, and optimizing daily traffic timing \cite{gu2021survey}.\\
Machine learning and deep learning models form the bedrock of human activity recognition systems\cite{islam2023multi}. Deep learning, in particular, has witnessed significant advancements in handling complex data across various computer vision domains \cite{dahou2022human}. Different classification techniques, such as the Markov model, decision tree, K-nearest neighbor (KNN), and artificial neural network (ANN), have been utilized for human activity recognition\cite{shen2023federated}. Notably, deep learning models have excelled in reducing preprocessing steps while facilitating automatic and dynamic feature extraction \cite{pouyanfar2018survey}. Their success extends to extracting valuable features from diverse raw data sources, which, along with using relevant data through humans and devices, has led to high evaluation accuracy \cite{gu2017locomotion}.\\
This research introduces IHARDS-CNN, a novel approach that combines various integrated datasets. It utilizes a one-dimensional CNN for human activity recognition (HAR). The amalgamation of three datasets, namely UCI-HAR, KU-HAR, and WISDM, form the foundation of the new dataset. One substantial advantage of integrating these datasets is enhancing speed by eliminating multiple additional algorithmic steps. Figure 1 illustrates the primary challenge addressed in human activity recognition through wearable sensors in this research. Consequently, this research poses two questions:

%Fig1

\begin{wrapfigure}{l}{0.5\textwidth}
\includegraphics[width=0.5\textwidth, keepaspectratio]{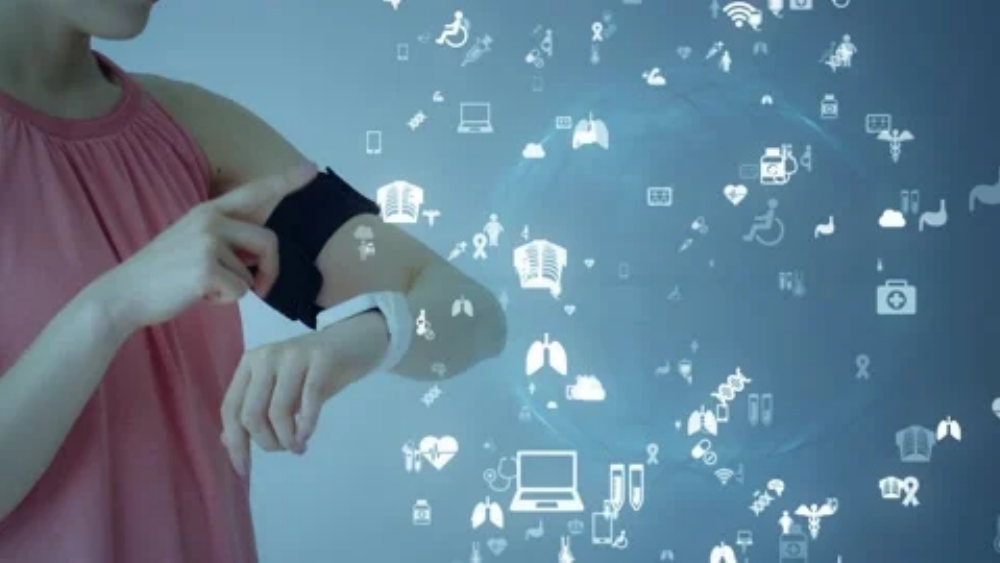}
\captionsetup{width=0.5\textwidth}
\caption{human activity recognition with smart watch\label{fig:figure1}}
\end{wrapfigure} 

RQ1: Can deep learning algorithms extract high-level features during the early stages of biometric segmentation?

RQ2: Does classification based on the high-level features extracted from the integrated data exhibit accuracy?

In this study, the innovation centers on aggregating biometric data and utilizing deep convolutional neural networks to extract combined patterns and classifications based on desired patterns. This approach obviates the need for secondary aggregation and achieves the desired results without encountering classification errors.

This research consists of six parts. The second part examines the research background. In the third section, we explain the proposed approach. The fourth section outlines related methods. In the fifth section, we report experimental results. The sixth section contains the conclusion and future works.

\section{Research Background}
In 2018, Zhu and his colleagues presented a semi-supervised approach to deep learning and feature extraction using Distributed Long Short-Term Memory (DLSTM) for Human Activity Recognition (HAR). In this approach, the authors used the UCI-HAR dataset. The proposed DLSTM, with sequential data modeling and high-level learning, can use either labeled or unlabeled data. However, it cannot detect unseen classes \cite{zhu2018novel}.

In 2018, Hassan and his colleagues introduced an approach using accelerometers and GYROSCOP sensors to reduce extraction dimensions. They employed a Deep Belief Network (DBN) for training in this method. Additionally, they used KPCA and strong feature sensors to reduce dimensions for signal extraction. Comparisons with multi-class SVM and other techniques demonstrated superior performance. However, this method needs more efficiency in real-time or complex situations \cite{hassan2018robust}.

In 2019, Jiang and Yong presented a WiFi CSI-based approach utilizing a Hybrid Attention mechanism-based Residual Neural Network (HARNN) with four detection techniques. They employed the linear regression method to search for optimal parameters. Additionally, two features, CPV and TFA, were used as extracted statistical parameters in the RNN. A two-layered decision tree was employed to identify activities after undergoing variance and correlation coefficient steps and removing random noise. The significance of this framework lies in its comparisons in two different environments and the subsequent comparison of results \cite{ding2019wifi}.

In 2020, Zhang and his colleagues introduced a Wi-Fi-based human activity recognition (HAR) system that combines eight CSI methods. This system employed three groups of activity data synthesis and eight types of conversion methods, including time stretching, Gaussian noise, frequency filter, etc. This system successfully addresses the challenge of mitigating the impact of small data on processing \cite{zhang2020data}.

In 2020, Zhou and his colleagues designed a semi-supervised deep learning framework based on training classifications from weakly labeled data and analyzing them. In this design, they employed the DQN technique for intelligent automatic tagging to resolve inappropriate tags. They applied an automatic label module and LSTM to this design. They used LSTM techniques to handle sequential motion data. Furthermore, this design developed a mechanism to integrate data from multiple sensors for accurate identification \cite{zhou2020deep}.

In 2020, Yong et al. presented a feature selection method for HAR based on the Pearson correlation coefficient. This method used three deep learning classification techniques for evaluation: Random Forest, Naïve Bayes, and C4.5. It utilized two separate datasets published by Washington State University, each providing diverse datasets. The feature selection process consisted of three steps: first, they converted a vector into the feature of the correlation coefficient. Then, they evaluated the relationship between the features by the weighting coefficient, and finally, they deleted the remaining related features. This method detects activities and is particularly significant regarding sensor facilitation in smart homes, energy management, and assistance for Alzheimer's patients \cite{liu2020daily}.

In 2020, Katrina and her colleagues implemented four different scenarios to increase the efficiency of activity recognition. This method employed LOSOCV as a reasonable criterion despite the need to repeat the training and testing process for evaluation. This method utilizes the MHEALTH dataset. Two deep learning models, FFNNs and CNN, were employed for better activity recognition. However, this method has the disadvantage of removing the signal sign and increasing the training time\cite{gholamiangonabadi2020deep}.

In 2020, Mohammad and his colleagues introduced a new framework called MS-IE HHAR for heterogeneous sensors in the Internet of Things environment. This model comprised two stages: the first stage involved time series imaging, and the second stage consisted of various deep learning models. The two channels of the model included the first channel with a multi-headed convolutional module and the second channel comprising six convolutional layers. The UCI-HAR and MEHEALTH datasets were used for evaluation, respectively\cite{abdel2020deep}.

In 2021, Sakron and Anochit presented a deep-learning network detection model based on LSTM. In this model, they improved the accuracy rate by approximately 2\% using the four-layered base network CNN-LSTM. The dataset used in this model is UCI-HAR. They utilized a Bayesian optimizer to set the meta-parameters of each LSTM. The advantage of this model is the LSTM layers that improve the extracted features regarding time dependency. They tested five LSTM-based models using validation data and the training meta-parameters approach\cite{mekruksavanich2021lstm}.

In 2021, Javed and his colleagues used a Deep Recurrent Neural Network (DRNN) for data from embedded hardware sensors. They collected data to detect combined activity with the GOOGLE FIT API. They introduced a systematic and practical framework for continuous activities by examining different DRNN parameters for monitoring. The framework evaluated in this research used both ABA and WABA scenarios. However, this model's credibility suffers from the low number of participants in different age groups\cite{javed2021smartphone}.

In 2022, xia et al. presented a deep neural network with LSTM for multi-parameter automatic extraction and classification. In this model, they first entered the raw data from the sensor into convolution and then into two layers of LSTM. Next, they used the integration layer (GAP) to replace the connection layer to reduce the parameter. The advantage of this model is the high detection capability despite fewer parameters. The datasets used to evaluate this model are UCI-HAR, WISDM, and OPPORTUNITY\cite{xia2020lstm}.

Based on previous methods and studies, Table 1 summarizes their advantages and disadvantages.

Various types of Human Activity Recognition (HAR) datasets are available, with the most famous compiled by KAIXUANCHEN et al \cite{chen2021deep}. In Table No. 2, the coding of the type of classes and Table 3 (with the titles of count of participants, type of activity, and challenges) briefly display this information.
%table 1,2,3
\begin{table}[h]
\centering
\caption{A side-by-side comparison of the mentioned methods}
\label{tab:table1}
\resizebox{\columnwidth}{!}{%
\begin{tabular}{|c|c|c|c|c|}
\hline \xrowht{18pt}
\textbf{AUTHOR}                                                        & \textbf{Approach}                                                                                                                   & \textbf{Advantages}                                                                                                                                              & \textbf{Disadvantages}                                                                                                                                   & \textbf{Dataset}                                                      \\ \hline
\begin{tabular}[c]{@{}c@{}}Zhang\\ et al\cite{zhang2020data}\end{tabular}          & \begin{tabular}[c]{@{}c@{}}Proposing a system\\ based on WIFI\\ combined with CSI\end{tabular}                                      & \begin{tabular}[c]{@{}c@{}}1.Disable small data\\ 2.Dense-LSTM model\\ increases accuracy by\\ up to 21.2\% compared\\ to the traditional DNN model\end{tabular} & \begin{tabular}[c]{@{}c@{}}limited set of data\\ and fails to account\\ for future unseen data\end{tabular}                                              & Others*                                                               \\ \hline 
\begin{tabular}[c]{@{}c@{}}Xiaokang\\ et al\cite{zhou2020deep}\end{tabular}       & \begin{tabular}[c]{@{}c@{}}Proposing a framework\\ with DQN technique for\\ automatic labeling\end{tabular}                         & \begin{tabular}[c]{@{}c@{}}1.Continuous monitoring\\  with the Internet of Things\\ 2.Multi-sensor data fusion\end{tabular}                                      & \begin{tabular}[c]{@{}c@{}}1.The need for infrastructure\\ equipment for implementation\\ 2.Time consuming data \\ labeling during training\end{tabular} & Others*                                                               \\ \hline 
\begin{tabular}[c]{@{}c@{}}Sakorn \&\\ Anuchit\cite{mekruksavanich2021lstm}\end{tabular}    & Proposing a CNN-LSTM                                                                                                                & \begin{tabular}[c]{@{}c@{}}Increased accuracy Improved\\ extraction features\end{tabular}                                                                        & \begin{tabular}[c]{@{}c@{}}Complexity in \\ implementation\end{tabular}                                                                                  & UCI-HAR                                                               \\ \hline 
\begin{tabular}[c]{@{}c@{}}XIA\\ et al\cite{xia2020lstm}\end{tabular}            & \begin{tabular}[c]{@{}c@{}}Proposing a deep learning\\ LSTM with RNN\end{tabular}                                                   & \begin{tabular}[c]{@{}c@{}}Reduced model parameters\\ with GAP layer\end{tabular}                                                                                & \begin{tabular}[c]{@{}c@{}}High computational\\ cost\end{tabular}                                                                                        & \begin{tabular}[c]{@{}c@{}}UCI-HAR\\ WISDM\\ OPPURTUNITY\end{tabular} \\ \hline 
\begin{tabular}[c]{@{}c@{}}Javed\\ et al\cite{javed2021smartphone}\end{tabular}          & \begin{tabular}[c]{@{}c@{}}Proposing deep learning\\  based on DRNN\end{tabular}                                                    & \begin{tabular}[c]{@{}c@{}}Improved at identifying\\  complex activities\end{tabular}                                                                            & \begin{tabular}[c]{@{}c@{}}Limited number of\\ participants for\\ evaluation\\     Limitation of privacy\end{tabular}                                    & Others*                                                               \\ \hline 
\begin{tabular}[c]{@{}c@{}}JIANYANG \\ \& YONG\\ \cite{ding2019wifi}\end{tabular} & \begin{tabular}[c]{@{}c@{}}Proposing a WIFI CISI\\ with HRRNN\end{tabular}                                                          & \begin{tabular}[c]{@{}c@{}}1.wide coverage\\ 2.cost effective\end{tabular}                                                                                       & \begin{tabular}[c]{@{}c@{}}1.privacy concern\\ 2.envirnomental \\ interference\end{tabular}                                                              & Others*                                                               \\ \hline 
\begin{tabular}[c]{@{}c@{}}Yong\\ et al\cite{liu2020daily}\end{tabular}           & \begin{tabular}[c]{@{}c@{}}Proposing 3 deep learning\\ classification techniques: \\ NAVE BAYES, RANDOM\\ FOREST, C405\end{tabular} & \begin{tabular}[c]{@{}c@{}}Improved assistance in taking\\ care of the elderly, including\\ Alzheimer's patients\\ Improved feature selection\end{tabular}       & \begin{tabular}[c]{@{}c@{}}May not be suitable \\ for all types of smart\\ homes\end{tabular}                                                            & Others*                                                               \\ \hline 
\begin{tabular}[c]{@{}c@{}}KATARINA\\ et al\cite{gholamiangonabadi2020deep}\end{tabular}        & \begin{tabular}[c]{@{}c@{}}Proposing a LOSOV with\\ no repetitions\end{tabular}                                                     & \begin{tabular}[c]{@{}c@{}}Independent estimate of the \\ performance for new subject\end{tabular}                                                               & \begin{tabular}[c]{@{}c@{}}Increased time\\ of training\end{tabular}                                                                                     & MHEALTH                                                               \\ \hline 
\begin{tabular}[c]{@{}c@{}}Zhu et al\\ \cite{zhu2018novel}\end{tabular}           & Proposing a DLSTM                                                                                                                   & High-level feature extraction                                                                                                                                    & \begin{tabular}[c]{@{}c@{}}Unable to detect \\ all classes\end{tabular}                                                                                  & UCI-HAR                                                               \\ \hline 
\begin{tabular}[c]{@{}c@{}}Hassan\\ et al\cite{hassan2018robust}\end{tabular}          & \begin{tabular}[c]{@{}c@{}}Proposing a GYROSCOP\\ sensor for decreasing\\ dimensions\end{tabular}                                   & Solves the privacy problem                                                                                                                                       & \begin{tabular}[c]{@{}c@{}}Complexity\\ Problem identifying \\ complex activities\end{tabular}                                                           & UCI-HAR                                                               \\ \hline 
\begin{tabular}[c]{@{}c@{}}Basset\\ et al\cite{abdel2020deep}\end{tabular}         & \begin{tabular}[c]{@{}c@{}}Proposing a new framework\\  named MS-IE HHAR from\\  heterogeneous sensors\end{tabular}                 & \begin{tabular}[c]{@{}c@{}}1.Runs with minimal computing \\ resources 2.Processing of various\\  sensor data in IoT applications\end{tabular}                    & \begin{tabular}[c]{@{}c@{}}Needs a lot of \\ data for training\end{tabular}                                                                              & \begin{tabular}[c]{@{}c@{}}UCI-HHAR\\ MEHEALTH\end{tabular}           \\ \hline
\end{tabular}%
}
\end{table}

%table 2
\begin{table}[h]
\centering
\caption{Class Code}
\label{tab:table2}
\resizebox{\columnwidth}{!}{%
\begin{tabular}{|c|c|c|c|c|c|}
\hline \xrowht{18pt}
\textbf{Code} & \textbf{Class Name} & \textbf{Code} & \textbf{Class Name} & \textbf{Code} & \textbf{Class Name} \\ \hline
1 & Standing & 45 & eating chips & 89 & Turning off light \\ \hline \xrowht{12pt}
2 & Walking & 46 & eating sandwich & 90 & Preparing breakfast \\ \hline \xrowht{12pt}
3 & Sitting & 47 & eating yogurt & 91 & eating breakfast \\ \hline\xrowht{12pt}
4 & Lying on back & 48 & pouring water from a bottle to a glass & 92 & making coffee \\ \hline\xrowht{12pt}
5 & Lying on right side & 49 & opening refrigerator & 93 & drinking coffee \\ \hline\xrowht{12pt}
6 & Ascending stairs & 50 & taking out a container from refrigerator & 94 & listing to music \\ \hline\xrowht{12pt}
7 & Descending stairs & 51 & Putting back a container in to refrigerator & 95 & dusting furniture \\ \hline\xrowht{12pt}
8 & Standing up & 52 & opening cupboard & 96 & sweeping floor \\ \hline\xrowht{12pt}
9 & Sitting down & 53 & closing cupboard & 97 & mopping floor \\ \hline\xrowht{12pt}
10 & Jogging & 54 & taking out a plate & 98 & watering plants \\ \hline\xrowht{12pt}
11 & Frezing of gait & 55 & using a microwave & 99 & feeding pets \\ \hline\xrowht{12pt}
12 & Running & 56 & using oven & 100 & leaving home \\ \hline\xrowht{12pt}
13 & Jumping & 57 & washing dishes & 101 & entering home \\ \hline\xrowht{12pt}
14 & boxing & 58 & eating & 102 & doing laundry \\ \hline\xrowht{12pt}
15 & waving & 59 & standing and relaxing & 103 & making bed \\ \hline\xrowht{12pt}
16 & watching TV & 60 & waist bends forward & 104 & reading book \\ \hline\xrowht{12pt}
17 & house cleaning & 61 & frontal elevation of arms & 105 & writing in journal \\ \hline
18 & cycling & 62 & knees bending(crouching) & 106 & meditating \\ \hline\xrowht{12pt}
19 & Nordic walking & 63 & normal activity & 107 & doing yoga \\ \hline\xrowht{12pt}
20 & computer work & 64 & Heart failure episode & 108 & exercising \\ \hline\xrowht{12pt}
21 & car driving & 65 & walking while carrying on object & 109 & taking medication \\ \hline\xrowht{12pt}
22 & vacum cleaning & 66 & walking while pushing on object & 110 & brushing hair \\ \hline\xrowht{12pt}
23 & ironing & 67 & walking while pulling on object & 111 & applying makeup \\ \hline\xrowht{12pt}
24 & folding laundry & 68 & lying on stomach & 112 & hand clapping \\ \hline\xrowht{12pt}
25 & playing soccer & 69 & lying on left side & 113 & backward fall \\ \hline\xrowht{12pt}
26 & rope jumping & 70 & sitting to standing & 114 & fall while lying \\ \hline\xrowht{12pt}
27 & standing still & 71 & standing to sitting & 115 & falling \\ \hline\xrowht{12pt}
28 & walking at normal speed & 72 & standing to lying & 116 & aerobics \\ \hline\xrowht{12pt}
29 & walking at fast speed & 73 & lying to standing & 117 & elliptical training \\ \hline\xrowht{12pt}
30 & sitting on chair & 74 & standing to lying & 118 & rowing \\ \hline\xrowht{12pt}
31 & sitting on bed & 75 & sitting to walking & 119 & stair stepper training \\ \hline\xrowht{12pt}
32 & standing up from chair & 76 & sleeping & 120 & talk-sit \\ \hline\xrowht{12pt}
33 & standing up from bed & 77 & Bed to- chair transfer & 121 & talk-stand \\ \hline\xrowht{12pt}
34 & lying down on bed from sitting position & 78 & Chair-to-bed transfer & 122 & walk-backward \\ \hline\xrowht{12pt}
35 & sitting down on chair from standing position & 79 & toilet use & 123 & walk-circle \\ \hline\xrowht{12pt}
36 & moving in a wheel chair & 80 & Shower/ Bath & 124 & table-tennis \\ \hline\xrowht{12pt}
37 & standing on one leg with eyes open & 81 & dressing & 125 & pick \\ \hline\xrowht{12pt}
38 & standing on one leg with eyes closed & 82 & grooming & 126 & Lay-stand \\ \hline\xrowht{12pt}
39 & standing on tiptoes & 83 & Forward fall & 127 & sit-up \\ \hline\xrowht{12pt}
40 & stretching arms & 84 & sideway s fall & 128 & push up \\ \hline\xrowht{12pt}
41 & nose touching & 85 & Backward fall & 129 & squatting \\ \hline\xrowht{12pt}
42 & drinking water & 86 & stumbling & 130 & Stair ascent and descent \\ \hline\xrowht{12pt}
43 & brushing teeth & 87 & taking shower & 131 & face washing \\ \hline\xrowht{12pt}
44 & eating apple & 88 & drying off with towel & 132 & Using mobile phone \\ \hline
\end{tabular}%
}
\end{table}

%table 3
\begin{landscape}
\begin{longtable}[c]{|c|c|c|c|c|c|c|c|}
\caption{Different types of dataset available in the field of human activity detection}

\label{tab:table3}\\

\hline
\xrowht{18pt}
\textbf{Number} & \textbf{Dataset}                                                            & \textbf{Context}                                                                  & \textbf{Subject} & \textbf{Activities}                                            & \textbf{Class Names}                                                                                                                                                      & \textbf{Sensor types}                                                 & \textbf{Challenges}                                                        \\ \hline
\endfirsthead

\multicolumn{8}{c}
{{\bfseries \tablename \thetable{} -- Different types of dataset available in the field of human activity detection(continue)}}\\
\hline\xrowht{18pt}
\textbf{Number} & \textbf{Dataset}                                                            & \textbf{Context}                                                                  & \textbf{Subject} & \textbf{Activities}                                            & \textbf{Class Names}                                                                                                                                                      & \textbf{Sensor types}                                                 & \textbf{Challenges}                                                        \\ \hline

\endhead 
1               & \begin{tabular}[c]{@{}c@{}}WISDM\\ activity\\ predictions\end{tabular}      & Daily Living                                                                      & 29               & 6                                                              & 1, 2,3, 6,7                                                                                                                                                               & Wearable                                                              & \begin{tabular}[c]{@{}c@{}}Class\\ Imbalance\end{tabular}                  \\ \hline 
2               & UCI-HAR                                                                     & Daily Living                                                                      & 30               & 6                                                              & 1,2,3,4,6,7                                                                                                                                                               & Wearable                                                              & Multimodal                                                                 \\ \hline 
3               & opportunity                                                                 & Daily Living                                                                      & 4                & 9                                                              & 1,2,3,4,5,6,7,8,9                                                                                                                                                         & \begin{tabular}[c]{@{}c@{}}Wearable,\\ Object,\\ Ambient\end{tabular} & \begin{tabular}[c]{@{}c@{}}Multimodal,\\ Composite\\ Activity\end{tabular} \\ \hline 
4               & \begin{tabular}[c]{@{}c@{}}Daphnet\\ Freezing\\ Of Gait\end{tabular}        & \begin{tabular}[c]{@{}c@{}}Patients of \\ Parkinson s\\ Disease\end{tabular}      & 10               & 3                                                              & 1,2,11                                                                                                                                                                    & Wearable                                                              & Simple                                                                     \\ \hline 
5               & \begin{tabular}[c]{@{}c@{}}Berkeley\\ MHAD\end{tabular}                     & Daily Living                                                                      & 12               & 11                                                             & \begin{tabular}[c]{@{}c@{}}1,2,3,4,6,7,10,12,\\ 13,14,15\end{tabular}                                                                                                     & \begin{tabular}[c]{@{}c@{}}Wearable,\\ Ambient\end{tabular}           & Multimodal                                                                 \\ \hline 
6               & PAMAP2                                                                      & Daily Living                                                                      & 9                & 18                                                             & \begin{tabular}[c]{@{}c@{}}1,2,3,4,6,7,12,16,\\ 17,18,19,20,21,\\ 22,23,24,25,26\end{tabular}                                                                             & Wearable                                                              & Multimodal                                                                 \\ \hline 
7               & SHO                                                                         & Daily Living                                                                      & 10               & 7                                                              & 1,2,3,4,12,13,129                                                                                                                                                         & Wearable                                                              & Simple                                                                     \\ \hline 
8               & UCI-HAPT                                                                    & \begin{tabular}[c]{@{}c@{}}Daily Living\\ With activity\\ transition\end{tabular} & 30               & 6                                                              & 1,2,3,4,7,10                                                                                                                                                              & Wearable                                                              & Multimodal                                                                 \\ \hline 
9               & HHAR                                                                        & Daily Living                                                                      & 9                & 6                                                              & 2,3,4,6,12,27                                                                                                                                                             & Wearable                                                              & Multimodal                                                                 \\ \hline 
10              & ARAS                                                                        & \begin{tabular}[c]{@{}c@{}}Real-World\\ Home Living\end{tabular}                  & 2                & 27                                                             & \begin{tabular}[c]{@{}c@{}}10,12,13,27,28,29,\\ 30,31,33,34,35,36,\\ 37,38,39,40,41,42,\\ 43,44,45,46,47,48\end{tabular}                                                  & \begin{tabular}[c]{@{}c@{}}Ambient,\\ Object\end{tabular}             & Multimodal                                                                 \\ \hline 
11              & \begin{tabular}[c]{@{}c@{}}Ambient\\ Kitchen\end{tabular}                   & \begin{tabular}[c]{@{}c@{}}Food \\ Preparation\end{tabular}                       & 20               & 11                                                             & \begin{tabular}[c]{@{}c@{}}48,49,50,51,52,53,\\ 5,4,55,56,57\end{tabular}                                                                                                 & Object                                                                & Simple                                                                     \\ \hline 
12              & USC-HAD                                                                     & Daily Living                                                                      & 14               & 12                                                             & \begin{tabular}[c]{@{}c@{}}1,2,3,4,6,7,10,\\ 13,20,42,43,58,\end{tabular}                                                                                                 & Wearable                                                              & Multimodal                                                                 \\ \hline 
13              & MHEALTH                                                                     & \begin{tabular}[c]{@{}c@{}}Real-World\\ Home Living\end{tabular}                  & 10               & 12                                                             & \begin{tabular}[c]{@{}c@{}}1,2,4,6,10,12,13,\\ 18,59,60,61,62\end{tabular}                                                                                                & Wearable                                                              & Multimodal                                                                 \\ \hline 
14              & \begin{tabular}[c]{@{}c@{}}BIDMC \\ Congestive\\ Heart Failure\end{tabular} & Hear failure                                                                      & 15               & 2                                                              & 63,64                                                                                                                                                                     & Wearable                                                              & \begin{tabular}[c]{@{}c@{}}Class\\ Imbalance\end{tabular}                  \\ \hline 
15              & DSADS                                                                       & \begin{tabular}[c]{@{}c@{}}Daily Living\\ and sports\end{tabular}                 & 8                & 19                                                             & \begin{tabular}[c]{@{}c@{}}1,2,3,4,5,6,7,9,65,\\ 66,67,68,69,70,71,\\ 72,73,74,75,131\end{tabular}                                                                        & Wearable                                                              & Multimodal                                                                 \\ \hline 
16              & CASAS-4                                                                     & \begin{tabular}[c]{@{}c@{}}Real-World \\ Home Living\end{tabular}                 & 2                & 15                                                             & \begin{tabular}[c]{@{}c@{}}1,2,3,4,58,65,\\ 66,67,76,77,78,\\ 79,80,81,82\end{tabular}                                                                                    & \begin{tabular}[c]{@{}c@{}}Object,\\ Ambient\end{tabular}             & Multimodal                                                                 \\ \hline 
17              & \begin{tabular}[c]{@{}c@{}}SmartWatch,\\ Notch,\\ Farseeing\end{tabular}    & \begin{tabular}[c]{@{}c@{}}Daily Living \\ \& \\ Fall Detection\end{tabular}      & 7                & \begin{tabular}[c]{@{}c@{}}4 ADL \\ \& \\ 4 Fall\end{tabular}  & 1,2,3,4,83,84,85,86                                                                                                                                                       & Wearable                                                              & \begin{tabular}[c]{@{}c@{}}Class\\ Imbalance\end{tabular}                  \\ \hline 
18              & \begin{tabular}[c]{@{}c@{}}Darmstadt \\ Daily\\  Routines\end{tabular}      & \begin{tabular}[c]{@{}c@{}}Real-World \\ Routines\end{tabular}                    & 1                & 35                                                             & \begin{tabular}[c]{@{}c@{}}16,20,22,23,24,43,\\ 57,87,88,89,90,91,\\ 92,93,94,95,96,\\ 97,98,99,100,\\ 101,102,103,104,\\ 105,106,107,108,\\ 109,110,111,132\end{tabular} & Wearable                                                              & \begin{tabular}[c]{@{}c@{}}Class\\ Imbalance\end{tabular}                  \\ \hline 
19              & MotionSense                                                                 & Daily Living                                                                      & 24               & 6                                                              & 1,2,3,6,7,10                                                                                                                                                              & Wearable                                                              & Simple                                                                     \\ \hline 
20              & \begin{tabular}[c]{@{}c@{}}MobiAct,\\ MobiFall\end{tabular}                 & \begin{tabular}[c]{@{}c@{}}Daily Living \\ \& \\ Fall Detection\end{tabular}      & 66               & \begin{tabular}[c]{@{}c@{}}12 ADL \\ \& \\ 4 Fall\end{tabular} & \begin{tabular}[c]{@{}c@{}}2,10,3,1,4,6,7,12,\\ 13,112,43,58,83,84,\\ 85,114\end{tabular}                                                                                 & Wearable                                                              & Multimodal                                                                 \\ \hline 
21              & \begin{tabular}[c]{@{}c@{}}Vankasteren\\ benchmark\end{tabular}             & \begin{tabular}[c]{@{}c@{}}Real-World \\ Home Living\end{tabular}                 & 3                & 9                                                              & 1,2,3,4,6,7,10,13,115                                                                                                                                                     & Object                                                                & Simple                                                                     \\ \hline 
22              & ActiveMiles                                                                 & \begin{tabular}[c]{@{}c@{}}Real-World \\ Routines\end{tabular}                    & 10               & 7                                                              & \begin{tabular}[c]{@{}c@{}}2,12,18,116,117,\\ 118,119\end{tabular}                                                                                                        & Wearable                                                              & Multimodal                                                                 \\ \hline 
23              & ActRecTut                                                                   & \begin{tabular}[c]{@{}c@{}}Hand Geasture \\ \& \\ Playing Tennis\end{tabular}     & 2                & 12                                                             & \begin{tabular}[c]{@{}c@{}}1,2, 3, 4, 6,7,10,12,\\ 13,18,70, 71\end{tabular}                                                                                              & Wearable                                                              & Multimodal                                                                 \\ \hline 
24              & KU-HAR                                                                      & Daily Living                                                                      & 90               & 18                                                             & \begin{tabular}[c]{@{}c@{}}1,2,3,4,6,7,12,71,\\ 120,121,122,123,\\ 124,125,126,127,128\end{tabular}                                                                       & Wearable                                                              & \begin{tabular}[c]{@{}c@{}}Class\\ Imbalance\end{tabular} \\ \hline
\end{longtable}
\end{landscape}

\section{Suggested Method}\label{sec3}

This research proposes a new integrated dataset called IHARDS by combining three datasets. Previous researchers typically used diverse datasets independently as input for multiple algorithms. Therefore, using these datasets resulted in data aggregation in the classification stage. But, in this research, we emphasize the necessity of data integrity by utilizing diverse datasets with common features. This approach reduces errors and avoids aggregating the results in the classification stage. Our proposed method employs raw data processing collected from smartphone and smartwatch accelerometer sensors.

Our proposed approach consists of two stages: dataset integration and a convolutional neural network. The overall technique presented in Figure 2 displays the first step, which involves integrating data collected from several datasets. The CNN extracts high-level features from the first stage into the second stage.

%fig2
\begin{figure}[h]
\centering
	\includegraphics[width=0.6\textwidth, keepaspectratio]{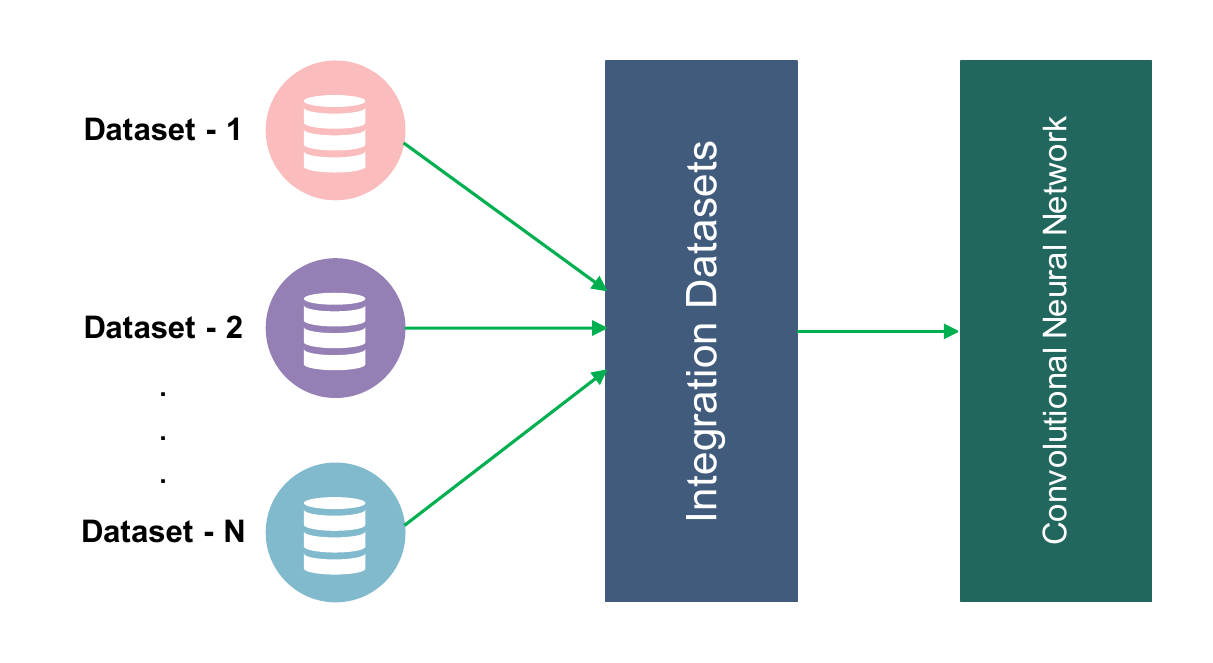}
	\captionsetup{width=0.5\textwidth}
	\caption{General architecture}
	\label{fig:figure2}
\end{figure}

\subsection{Proposed Integration Datasets \& Analysis}

This part explains the proposed dataset and the data generation and analysis method. In this phase, we used the UCI-HAR, WISDM, and KU-HAR datasets, and Table 4 displays the number of features along with their class names. Now, let us delve into the details of each dataset:

%table 4
\begin{table}[h]
\centering
\caption{Dataset Features}
\label{tab:table4}
\resizebox{\columnwidth}{!}{%
\begin{tabular}{|c|c|c|l|}
\hline\xrowht{15pt}
\textbf{Dataset Name} & \textbf{Number of Features} & \textbf{Number of Classes} & \multicolumn{1}{c|}{\textbf{Class Names}} \\ \hline\xrowht{20pt}
\textbf{KU-HAR\cite{sun2022human}} & 8 & 18 & \begin{tabular}[c]{@{}l@{}}1. Stand 2. Sit 3. Walk   4. Stair-down 5. Stair-up  6. Talk-sit   \\ 7. Talk-stand 8.Stand-sit  9. Lay 10.   Lay-stand 11. Pick \\ 12. Jump 13.Push-up 14. Sit-up  15. Walk-backwards \\ 16. Walk circle 17.   Run  18. Table-tennis\end{tabular} \\ \hline\xrowht{20pt}
\textbf{UCI-HAR\cite{fu2016human}} & 561 & 6 & 1. Stand 2. Sit 3.   Walk  4. Stair-down 5. Stair-up  6. Laying \\ \hline\xrowht{20pt}
\textbf{WISDM\cite{gholamiangonabadi2020deep}} & 6 & 6 & 1. Stand 2. Sit 3.   Walk  4. Stair-down 5. Stair-up  6. Jogging \\ \hline\xrowht{20pt}
\textbf{IHARDS-CNN Proposed} & 571 & 5 & 1.Stand 2.Sit 3.walk   4.Stair-down 5.Stair-up \\ \hline
\end{tabular}%
}
\end{table}

\subsubsection{UCI-HAR:}
This dataset was created by collecting data from body sensors with 30 participants. The data was collected while participants performed six daily behavioral and sports activities: standing, lying down, sitting, walking, going down, and going up. The dataset consists of 10,299 samples.
\subsubsection{WISDM:}
This dataset involved 36 participants performing daily activities, such as sitting, standing, jumping, climbing, and walking. This dataset’s aggregation of samples is 1,098,207. 
\subsubsection{KU-HAR:}
This dataset comprises 90 participants aged 18 to 34 and includes 18 types of daily activities. The dataset contains 1945 raw data samples.
\subsubsection{Proposed IHARDS:}
The new integrated dataset proposed in this research combines three previous datasets (UCI-HAR, KU-HAR, and WISDM), named Integrated Human Activity Recognition Dataset (IHARDS). This dataset contains five classes: walking, sitting, climbing stairs, descending stairs, and standing. And finally, five common classes are selected from the classes of three datasets. You can download the dataset from the following link: \href{https://www.kaggle.com/datasets/humanactivitydl/integrated-human-activity-recognition-dataset}{Download Dataset}

Initially, we followed the flowchart in Figure 3, where we stored each dataset's data separately in data frames and extracted five activity classes from each data frame. Next, we established three datasets, each with 420,000 random indices from each class, without repetition. After that, we read and compiled the ordered triple indices from the datasets into a single file with the name of the class. Finally, combining these five different class files, we constructed a new dataset with 2,100,000 instances.  

%fig3

\begin{figure}[h]
\centering
	\includegraphics[width=0.95\textwidth, keepaspectratio]{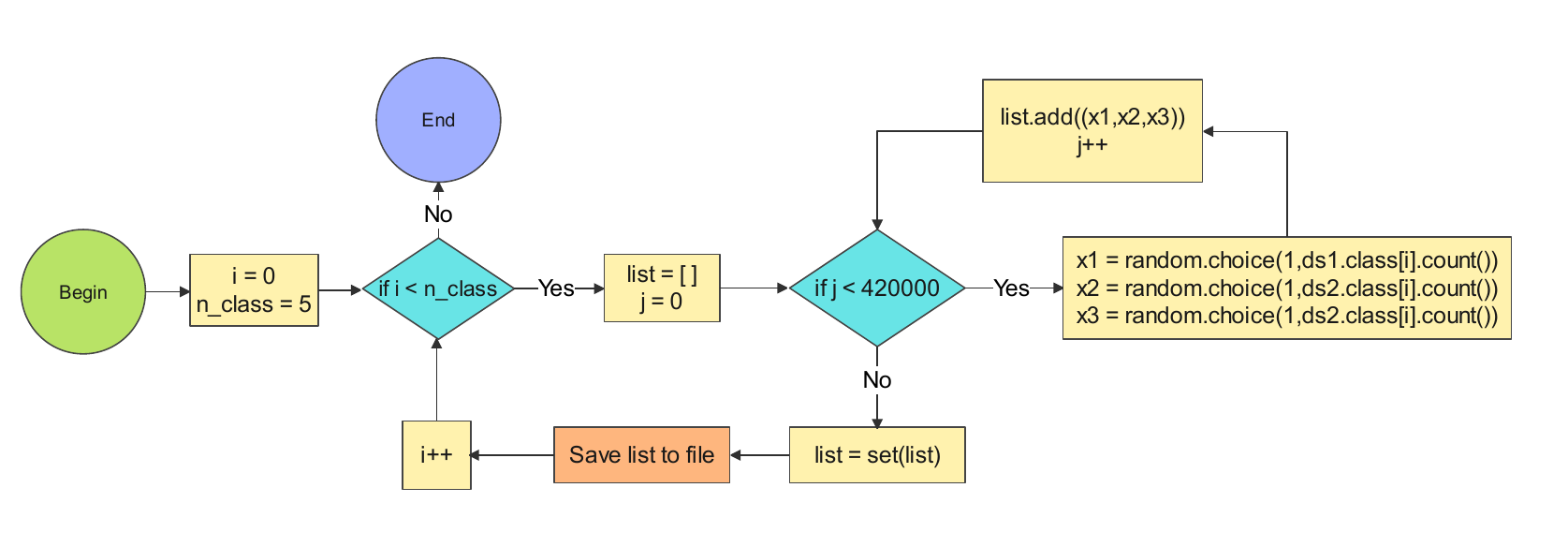}
	\captionsetup{width=0.5\textwidth}
	\caption{Integration dataset flowchart}
	\label{fig:figure3}
\end{figure}

\subsection{Analysis of the Proposed Dataset}
After creating the proposed dataset, we evaluated it in two models, with and without correlation, to check the correlation between features. In this research, we used Pearson's correlation coefficient to identify and remove correlated features to increase the accuracy of training, prediction, and processing speed.

The results of calculating the correlation coefficient revealed 460 attributes with a correlation coefficient above 50\% and 320 features with a correlation coefficient above 90\%. After removing those features from the dataset (reducing the dimensions), we were left with 111 features in IHARDS-0.5 and 251 features in IHARDS-0.9, which entered into a one-dimensional CNN as input for high-level feature extraction. Figure 4 displays a bar graph due to the limitation of illustrating features (571 dimensions). The process of extracting high-level features leads to the removal of noise and unnecessary features and the reduction of dimensions.

\begin{figure}[h]
\centering
	\includegraphics[width=0.75\textwidth, keepaspectratio]{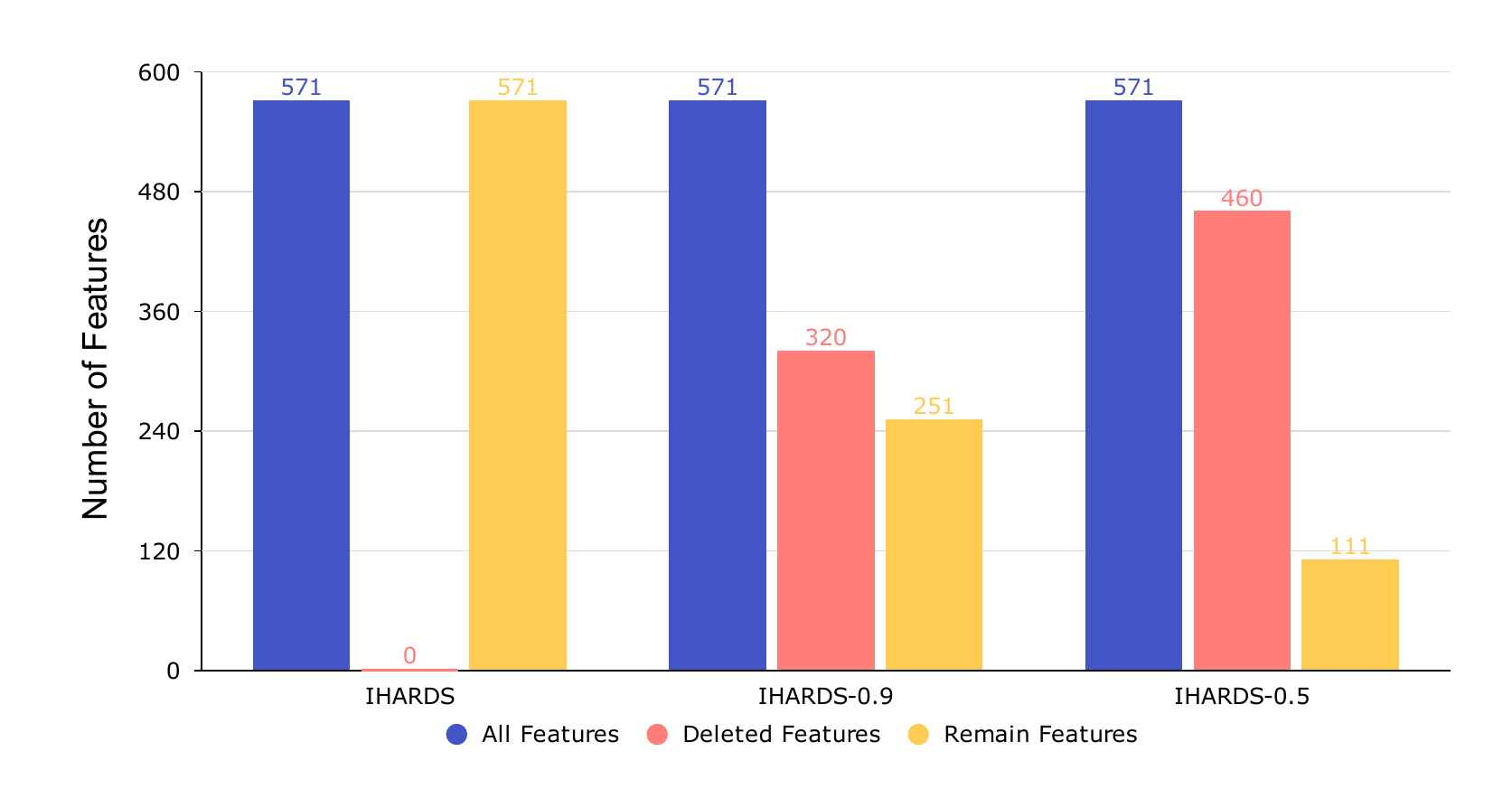}
	\captionsetup{width=0.50\textwidth}
	\caption{Number of datasets features}
	\label{fig:figure4}
\end{figure}

\subsubsection{DRWCC (Dimension Reduction with Calculating Correlation) Algorithm:}
Identifying and truncating the removable features from the IHARDS dataset after calculating the correlation coefficient is as follows:

\begin{figure}[h]
\centering
	\includegraphics[width=0.9\linewidth, keepaspectratio]{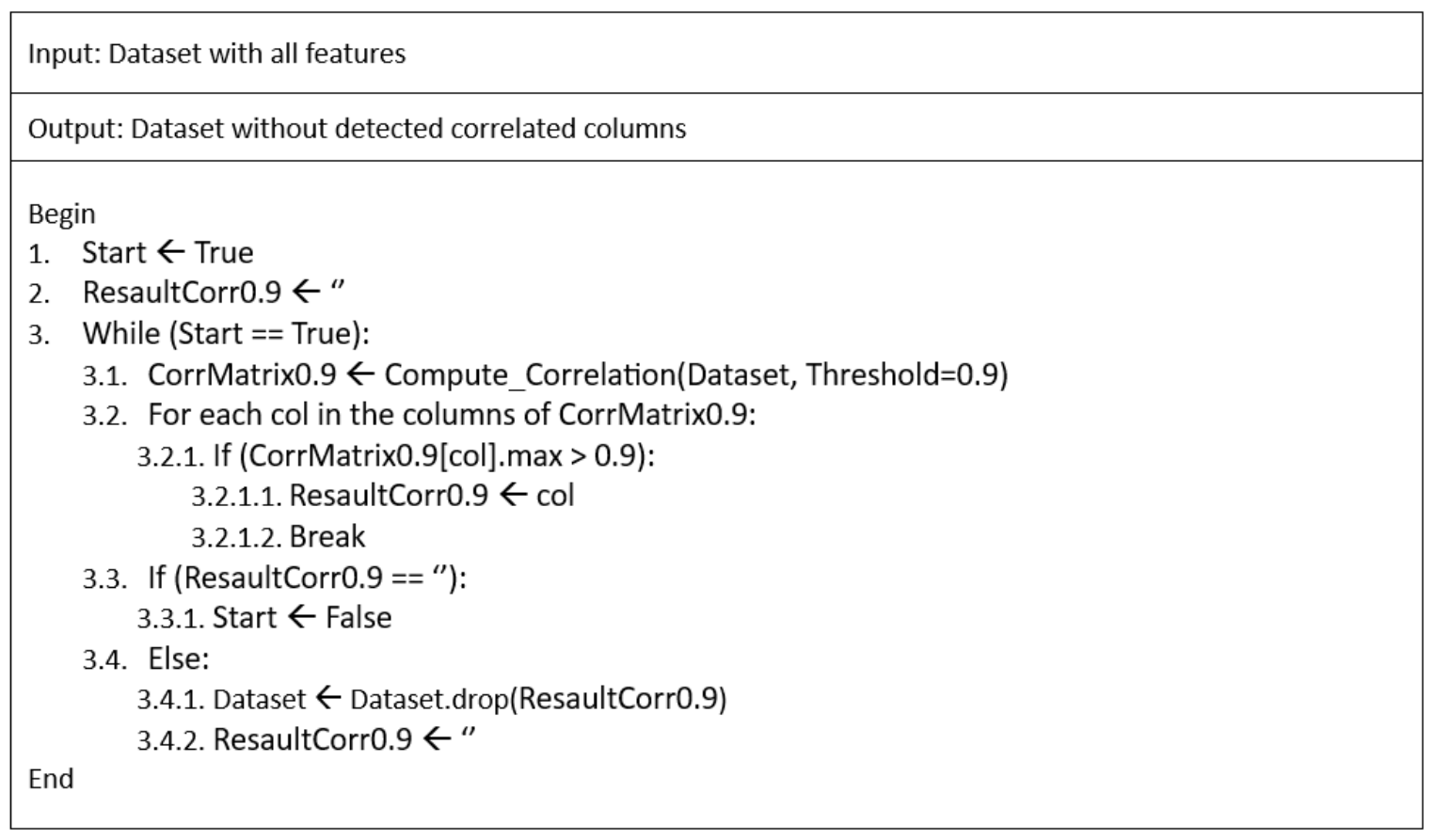}
	\label{fig:algorithm}
\end{figure}

\subsection{IDCNN:}
The Convolutional Neural Network (CNN) is a type of deep neural network model with a unique role in data processing, high-level feature extraction, and pattern recognition \cite{zhu2022continuous}. One of the essential features of CNN is using small filters from the data and applying them to the input data. The primary purpose of CNN is to extract high-level features.

Our proposed technique integrates the one-dimensional feature vector extracted from the CNN block in the fully connected layer to select the best feature. In the second step, the proposed method flattens the high-level features extracted from the CNN to enter the neural network. In the following steps, the extracted features enter the hidden layer of the neural network for training. We can employ regularity techniques such as batch normalization, dropout, and kernel constraint to enhance the stability and performance of the model. Figure 5 illustrates the steps of our proposed IHARDS-CNN approach for activity recognition.

\begin{figure}[h]
\centering
	\includegraphics[width=0.99\textwidth, keepaspectratio]{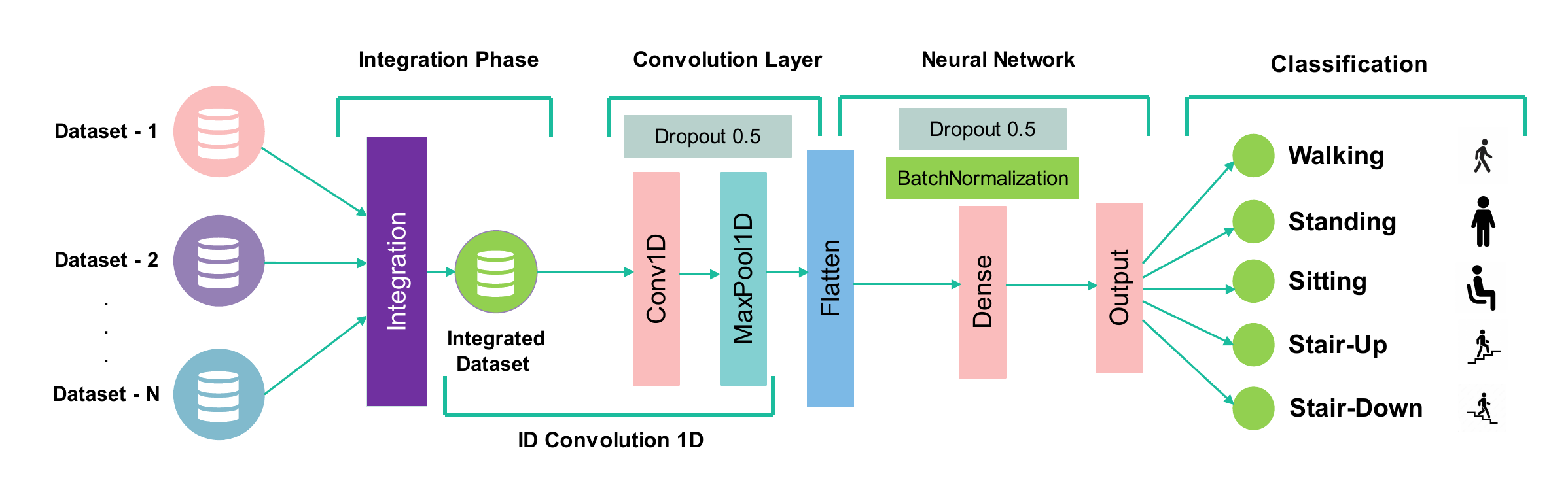}
	\captionsetup{width=0.5\textwidth}
	\caption{Proposed IHARDS-CNN Architecture}
	\label{fig:figure5}
\end{figure}

The Batch Normalization method makes the neural network faster and more stable by normalizing the input layers through re-centering and re-scaling\cite{garbin2020dropout}. Also, the Dropout method prevents over-fitting by randomly removing the weights in the neural network\cite{srivastava2014dropout}. As a result, not only is there no problem in using them simultaneously, but it also improves the model.

\section{Related Methods}
In this part, we briefly review several existing methods of CNN research in human activity recognition (HAR).

In 2019, Wan and his colleagues presented a smartphone accelerometer-based architecture and compared the advantages and disadvantages of five algorithms. This architecture includes preprocessing steps such as noise removal and normalization. Additionally, it utilizes a CNN to extract new features and evaluates different types of models (MLP, BLSTM, LSTM, SVM, and CNN) on the same dataset. The datasets used in this model are UCI-HAR and PAMP2, respectively\cite{wan2020deep}.

In 2020, Debadioti and his colleagues introduced the ENSEMCONVNET collection, which combines three models: ENCODED-NET, CNN-NET, and CNN-LSTM. These models consist of simple one-dimensional CNN cases that vary in the size of the kernel and the number of layers. In the decision-making phase, they use a combination of classification methods, such as majority voting. Currently, in cases with a high correlation rate between activities, this method achieves high accuracy in classification. This method uses datasets named WISDM, UNIMIB SHAR, and MOBIACT. For the final prediction, ENSEMCONVNET employs four standard hybrid techniques: majority voting, product sum law, and AWF4. Unfortunately, this method performs poorly in feature selection\cite{mukherjee2020ensemconvnet}.

In 2020, Cheng et al. presented a model called CONDCONV, based on CNN, aiming to increase capacity and efficiency in real-time mode. This model replaces the standard convolution layer with the linear combination of the CONDOCONV layer with the number of experts through gradient descent input training. Four datasets, including WISDM, PAMAP, UNIMIB-SHAR, and ORPORTUNITY, have been used for evaluation. Due to its simplicity of implementation and lack of need for preprocessing, this model applies the ideal division technique for signal time series to CNN input. CONDCONV is effective in HAR for real-time mobile applications and wearable devices due to its simplicity of implementation\cite{cheng2022real}.

In 2020, Seyed Wahab and Mohammad Javad introduced a new structure that combines LSTM’s short-term memory and network structures. This structure involves layers by skipping some to remove the gradient network effect left on the CNN. They evaluated the performance of this structure on the WISDM dataset, which contains signals recorded from six different activities, including walking, jumping, ascending, descending, sitting, and standing, and another dataset, including walking, jumping, sitting, standing, eating soup, eating sandwiches, and eating chips\cite{shojaedini2020mobile}.

In 2020, Mohamed and his colleagues introduced a supervised two-channeled model based on LSTM called the ST-DEEP HAR approach. This approach includes two learning paths for time representation, with a multi-channel CNN in each path. The first channel learns sequential data, while the second modifies compression. They trained and evaluated the model using two datasets, UCI-HAR and WISDM. One advantage of this two-channeled model is that it prevents the gradient effect from disappearing while emphasizing important related features\cite{abdel2020deep}.

In 2021, Nidhi and his colleagues presented a CNN-GRU multi-model-based model for human activity recognition (HAR). In this model, CNN extracts local features and records long-term dependencies in the data. Their model consists of an architecture with three different convolution sizes capable of recording different lengths of local correlations. To evaluate this model, they used three other datasets named UCI-HAR, WISDM, and PAMAP2. The advantage of this method lies in the combination of several models that can record various types of data and predict activity with less error\cite{dua2021multi}.

Helmi and colleagues in 2021 proposed an efficient system for HAR using feature selection called GBOGWO. The proposed algorithm combines gradient-based optimization (GBO) and gray wolf optimization (GWO) to overcome problems such as local optimality and slow convergence. This model achieved 98\% accuracy with SVM classification using the UCI-HAR and WISDM datasets\cite{helmi2021novel}.

In 2021, Tang and colleagues presented a lightweight CNN model based on LEGO filters. LEGO filters greatly reduce the number of parameters. In this model, the STE method is used to optimize the permutation of LEGO filters. Also, three-stage convolution (split - transform - merge) has been used to help cost reduction. Five datasets named UCI-HAR, WISDM, PAMAP2, UNIMIB-SHAR, and OPPORTUNITY have been used to evaluate HAR. The advantages of using the LEGO model are a great help in reducing the memory and cost of calculations and improving the performance despite the training error, in return for which the performance drops to some extent\cite{tang2020layer}.

In 2021, Zanobia and Jameel proposed an attention-based multi-head model for HAR. The framework of this model includes three lightweight convolutional heads, with each head using a one-dimensional CNN for feature extraction. In fact, the purpose of providing this simple deep learning architecture has been to be effective in extracting distinct features. The evaluation datasets used for this model are WISDM and UCI-HAR respectively. The advantage of this model is the suppression of unimportant items and automatic selection in the selection of features with the help of display ability. However, this method needs to determine the optimal size of the division to solve the weaknesses of the confusion matrix\cite{khan2021attention}.

In 2022, Denbo and his colleagues presented a new method called channel equalization to detect human activities by wearable sensors. This method is done with decorrelation on each mini-batch by activating all the different channels in the participation feature. This method has been evaluated in the Raspberry Pi platform Using this method without connecting raises the active channel to the network and does not depend on a specific channel, increasing the network's generalization. One of the challenges of this method is the network's weak performance\cite{huang2022channel}.

Shaik and Hossein, in 2023, used a genetic algorithm based on fuzzy to extract features from sensor data. The architecture used by them is DCNN-LSTM. Their goal is to combine this architecture with fuzzy recognition of complex behaviors. The datasets used in this research were UCI-HAR, PAMAP2, and WISDM. The advantage of combining this method with FGA is the compromise between accuracy and privacy. However, the data needs to be more balanced\cite{jameer2023dcnn}.

Table 5 briefly reviews the related works and their advantages and disadvantages.

%table 5
\captionof{table}{delivers a comparison of the related work.}
\resizebox{0.99\textwidth}{!}{%
\begin{tabular}{|ccccc|}
\hline
\multicolumn{5}{|c|}{\xrowht{20pt}\textbf{Human activity recognition}}                                                                                                                                                                                                                                                                                                                                                                                                                                                                                                                                                                                                                                                                                                 \\ \hline
\multicolumn{1}{|c|}{\xrowht{20pt}\textbf{Author}}                                                            & \multicolumn{1}{c|}{\xrowht{20pt}\textbf{dataset}}                                                                                & \multicolumn{1}{c|}{\xrowht{20pt}\textbf{Approach}}                                                                                                                       & \multicolumn{1}{c|}{\xrowht{20pt}\textbf{advantages}}                                                                                                                                                               & \xrowht{20pt}\textbf{disadvantages}                                                                                                                                  \\ \hline
\multicolumn{1}{|c|}{\begin{tabular}[c]{@{}c@{}}Basset et al\\ \cite{abdel2020deep}\end{tabular}}            & \multicolumn{1}{c|}{\begin{tabular}[c]{@{}c@{}}UCI-HAR,\\ WISDM\end{tabular}}                                        & \multicolumn{1}{c|}{Proposing a ST-DEEP HAR}                                                                                                                 & \multicolumn{1}{c|}{\begin{tabular}[c]{@{}c@{}}Prevents  the gradient from \\ disappearing\end{tabular}}                                                                                               & \begin{tabular}[c]{@{}c@{}}Not suitable in heterogeneous \\ environment\end{tabular}                                                                    \\ \hline 
\multicolumn{1}{|c|}{\begin{tabular}[c]{@{}c@{}}Nidhi et al\\ \cite{dua2021multi}\end{tabular}}             & \multicolumn{1}{c|}{\begin{tabular}[c]{@{}c@{}}UCI-HAR,\\ WISDM,\\ PAMAP2\end{tabular}}                              & \multicolumn{1}{c|}{Proposing   a CNN-GRU}                                                                                                                   & \multicolumn{1}{c|}{\begin{tabular}[c]{@{}c@{}}1.combines different data and \\ records it 2.Superior classification \\ performance 3.Automatic feature\\  extraction and classification\end{tabular}} & \begin{tabular}[c]{@{}c@{}}It suffers from\\ 1.Time consuming for large \\ labeled trained data  2.Performance \\ can be affected by noise\end{tabular} \\ \hline
\multicolumn{1}{|c|}{\begin{tabular}[c]{@{}c@{}}Helmi et al\\ \cite{helmi2021novel}\end{tabular}}             & \multicolumn{1}{c|}{\begin{tabular}[c]{@{}c@{}}UCI-HAR,\\ WISDM\end{tabular}}                                        & \multicolumn{1}{c|}{Proposing   a GBOGWO}                                                                                                                    & \multicolumn{1}{c|}{\begin{tabular}[c]{@{}c@{}}1.Reduced computational cost \\ 2.Increased interpretability\end{tabular}}                                                                              & \begin{tabular}[c]{@{}c@{}}Difficulty setting parameter \\ between  two algorithms\end{tabular}                                                         \\ \hline
\multicolumn{1}{|c|}{\begin{tabular}[c]{@{}c@{}}Wan et al\\ \cite{wan2020deep}\end{tabular}}               & \multicolumn{1}{c|}{\begin{tabular}[c]{@{}c@{}}UCI-HAR,\\ PAMP2\end{tabular}}                                        & \multicolumn{1}{c|}{\begin{tabular}[c]{@{}c@{}}Proposing an architecture based on\\ smartphone accelerometer\end{tabular}}                                   & \multicolumn{1}{c|}{Improved energy consumption cost}                                                                                                                                                  & It may not be accessible for everyone                                                                                                                   \\ \hline
\multicolumn{1}{|c|}{\begin{tabular}[c]{@{}c@{}}Tang et al\\ \cite{tang2020layer}\end{tabular}}              & \multicolumn{1}{c|}{\begin{tabular}[c]{@{}c@{}}UCI-HAR,\\ WISDM,\\ PAMAP2,\\ UNIMIB-SHAR\\ OPPORTUNITY\end{tabular}} & \multicolumn{1}{c|}{\begin{tabular}[c]{@{}c@{}}Proposing a  lightweight CNN model\\ based on LEGO filter\end{tabular}}                                       & \multicolumn{1}{c|}{Reduced memory consumption}                                                                                                                                                        & \begin{tabular}[c]{@{}c@{}}Limitations in the use of trained \\ networks  and requires real \\ data for evaluation\end{tabular}                         \\ \hline
\multicolumn{1}{|c|}{\begin{tabular}[c]{@{}c@{}}Mukherjee et al\\ \cite{mukherjee2020ensemconvnet}\end{tabular}}         & \multicolumn{1}{c|}{\begin{tabular}[c]{@{}c@{}}WISDM,\\ UNIMIB-SHAR\\ MOBIACT\end{tabular}}                          & \multicolumn{1}{c|}{\begin{tabular}[c]{@{}c@{}}Proposing a combination of three\\  models called ENSEMCONVNET\end{tabular}}                                  & \multicolumn{1}{c|}{Beneficial for healthcare}                                                                                                                                                         & Non-optimal feature selection                                                                                                                           \\ \hline
\multicolumn{1}{|c|}{\begin{tabular}[c]{@{}c@{}}Cheng et al\\ \cite{cheng2022real}\end{tabular}}             & \multicolumn{1}{c|}{\begin{tabular}[c]{@{}c@{}}WISDM,\\ PAMAP,\\ UNIMIB-SHAR \\ ORPORTUNITY\end{tabular}}            & \multicolumn{1}{c|}{\begin{tabular}[c]{@{}c@{}}Proposing a model called\\ CONDCONV \\ which is based on CNN\end{tabular}}                                    & \multicolumn{1}{c|}{\begin{tabular}[c]{@{}c@{}}Simplicity in implementation \\ in real-time use\end{tabular}}                                                                                          & \begin{tabular}[c]{@{}c@{}}Requires multivariate temporal sensor \\ data and computing power to run\end{tabular}                                        \\ \hline
\multicolumn{1}{|c|}{\begin{tabular}[c]{@{}c@{}}Shojaedini\\ \& Beirami\\ \cite{shojaedini2020mobile}\end{tabular}} & \multicolumn{1}{c|}{WISDM}                                                                                           & \multicolumn{1}{c|}{\begin{tabular}[c]{@{}c@{}}Proposing a new structure based \\ on a  combination of LSTM by\\  removing the gradient effect\end{tabular}} & \multicolumn{1}{c|}{\begin{tabular}[c]{@{}c@{}}1.Better detection of similar activities \\ 2.Removing the gradient effect\end{tabular}}                                                                & \begin{tabular}[c]{@{}c@{}}1.Complexity  2.The need \\ for large datasets for training\end{tabular}                                                     \\ \hline
\multicolumn{1}{|c|}{\begin{tabular}[c]{@{}c@{}}Zanobya\\ \& Jamil\\ \cite{khan2021attention}\end{tabular}}      & \multicolumn{1}{c|}{\begin{tabular}[c]{@{}c@{}}WISDM,\\ UCI-HAR\end{tabular}}                                        & \multicolumn{1}{c|}{\begin{tabular}[c]{@{}c@{}}Proposing An attention-based \\ multi-head model for HAR detection\end{tabular}}                              & \multicolumn{1}{c|}{\begin{tabular}[c]{@{}c@{}}Automatic selection in feature\\ selection\end{tabular}}                                                                                                & \begin{tabular}[c]{@{}c@{}}Low width and depth of convolution\\ head layers\end{tabular}                                                                \\ \hline

\multicolumn{1}{|c|}{\begin{tabular}[c]{@{}c@{}}denbo et al\\ \cite{huang2022channel}\end{tabular}}      & \multicolumn{1}{c|}{\begin{tabular}[c]{@{}c@{}}UCI-HAR, \\ WISDM,\\ PAMAP2, \\ UNIMIB-SHAR, \\ OPPORTUNITY, \\ USC-HAD\end{tabular}}                                        & \multicolumn{1}{c|}{\begin{tabular}[c]{@{}c@{}}Proposing a  Channel- \\ Equalization for HAR \\ detection\end{tabular}}                              & \multicolumn{1}{c|}{\begin{tabular}[c]{@{}c@{}}Increasing the network's \\ generalization\end{tabular}}                                                                                                & \begin{tabular}[c]{@{}c@{}}Weak performance\end{tabular}                                                                \\ \hline

\multicolumn{1}{|c|}{\begin{tabular}[c]{@{}c@{}}Shaik and Hossein\\ \cite{jameer2023dcnn}\end{tabular}}      & \multicolumn{1}{c|}{\begin{tabular}[c]{@{}c@{}}WISDM,\\ UCI-HAR, \\ PAMAP2\end{tabular}}                                        & \multicolumn{1}{c|}{\begin{tabular}[c]{@{}c@{}}Proposing a  DCNN-LSTM\end{tabular}}                              & \multicolumn{1}{c|}{\begin{tabular}[c]{@{}c@{}}Better detection complex \\ activities\end{tabular}}                                                                                                & \begin{tabular}[c]{@{}c@{}}need more data balance\end{tabular}                                                                \\ \hline

\end{tabular}%
}

\section{Experimental Results}
In this section, we discuss our proposed model's evaluation and its results. We review the results of the proposed dataset analysis in this section.

\subsection{Dataset:}
Table 6 displays the number of learning samples and their evaluation test results on our proposed dataset.

%table 6
\begin{table}[h]
\centering
\caption{Train and test samples split.}
\label{tab:table6}
\resizebox{0.5\textwidth}{!}{%
\begin{tabular}{c|c|}
\cline{2-2}\xrowht{15pt}
 & \textbf{IHARDS Proposed} \\ \hline
\multicolumn{1}{|c|}{Training samples} & 1050000 \\ \hline
\multicolumn{1}{|c|}{Testing samples} & 1050000 \\ \hline
\multicolumn{1}{|c|}{\textbf{Total}} & 2100000 \\ \hline
\end{tabular}%
}
\end{table}

\subsection{Parameters of the Proposed Model:}
We randomly use weight and bias during the learning phase in the simple one-dimensional CNN model. We employed the ADAM algorithm with a learning rate of 0.001, the $sparse\_categorical\_crossentropy$ loss function, and accuracy evaluation criteria as the improver in the proposed model. Additionally, this model demonstrates the desired result of evaluation accuracy before reaching overfitting. To evaluate the dataset, we performed ten iterations with an epoch value of ten and a batch size of 500. The results of this model have indicated high accuracy. Table 7 illustrates the model evaluation results on the dataset using evaluation criteria. According to the table, arch1 is more complex than the rest of the CNN architectures and fulfills the evaluation criteria by 100\%. Therefore, we attempted to maintain the percentage of evaluation criteria fulfillment in other architectures while simplifying the architecture. As a result, as you can see, Arch4 kept its 100\% accuracy and precision while being much more simplistic. Figure 6 depicts the error reduction process, and Figure 7 shows the increase in the accuracy of the CNN model during training. To create a unified dataset, we used a system with the following specifications:

\begin{itemize}
\item CPU: Intel Core i7 10870h.
\item GPU: Nvidia RTX 2070 8G.
\item Hard: SSD 1.5T Samsung.
\item RAM: 32G.
\end{itemize}

\begin{figure}[!h]
	\begin{minipage}[c]{0.5\linewidth}
		\begin{center}
		\includegraphics[width=1\textwidth, keepaspectratio]{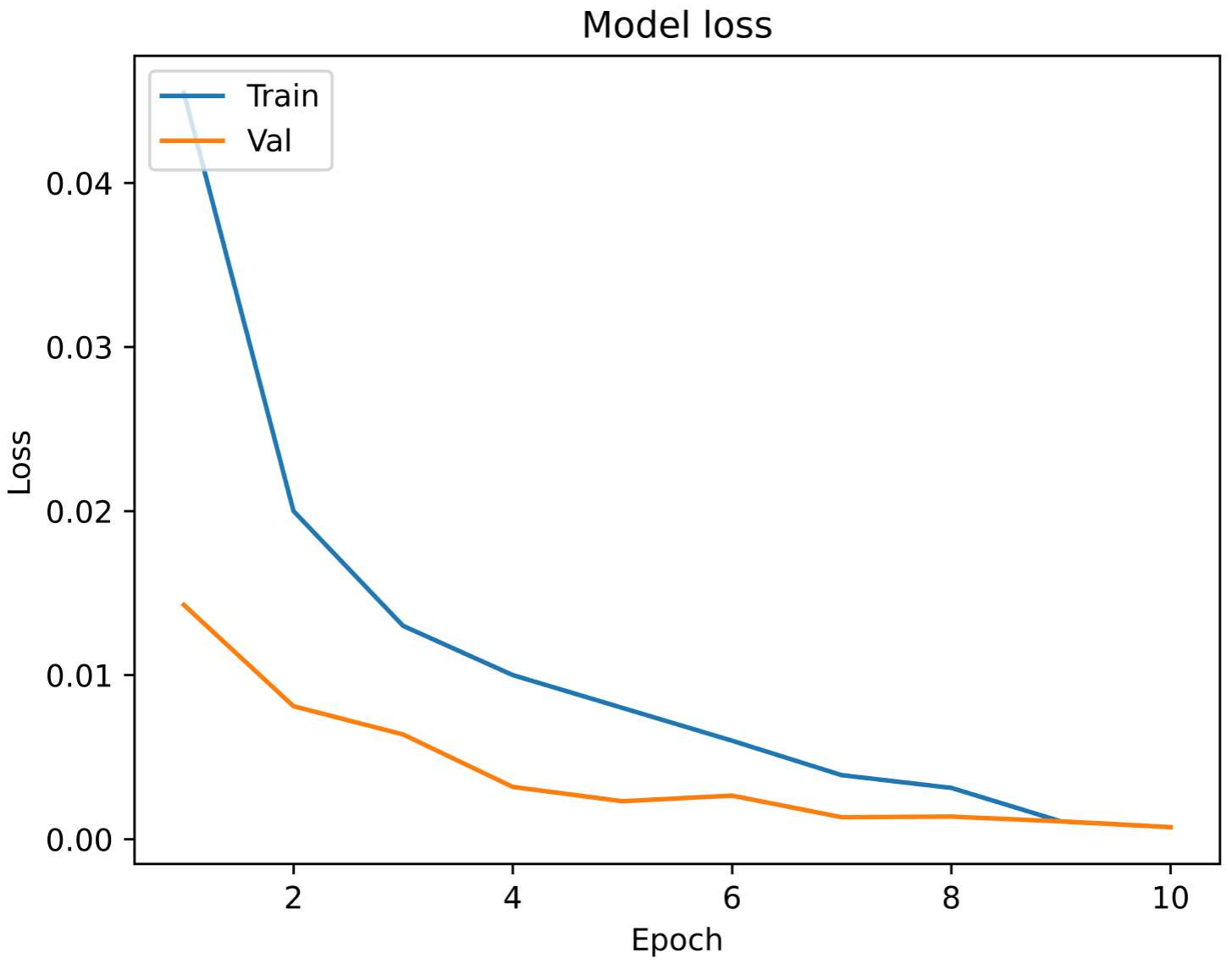}
		\caption{Error reduction}
		\end{center}
	\end{minipage}\hfill
	\begin{minipage}[c]{0.5\linewidth}
		\begin{center}
		\includegraphics[width=1\textwidth, keepaspectratio]{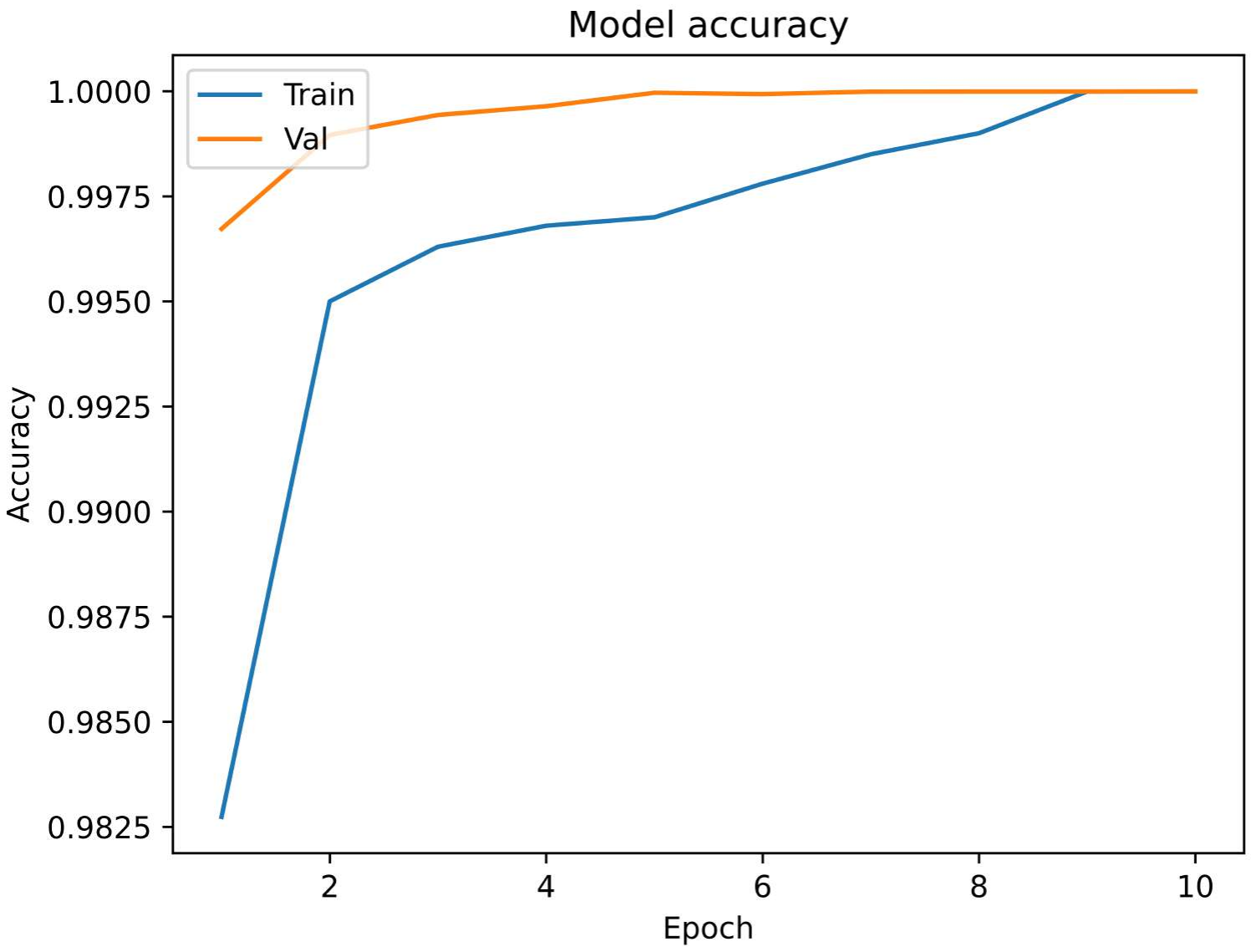}
		\caption{CNN accuracy training}
		\end{center}
	\end{minipage}
\end{figure}
  
To evaluate each architecture in terms of accuracy, recall, and f-measure, we calculated them using equations 1-4. In this calculation, TP represents true positive, TN is true negative, FP is false positive, and FN is false negative.

\begin{eqnarray}
SEN = \frac{TP}{TP+FN} \label{eq1}\\
SPF = \frac{TN}{TN+FP} \label{eq2}\\
ACC = \frac{TP+TN}{TP+TN+FP+FN} \label{eq3}\\
F1-measure(F1-m) = 2 * \frac{Prec * Rec}{Prec + Rec} \label{eq4}
\end{eqnarray}

Table 7 depicts the final results of the dataset evaluation. After removing the features with a high correlation coefficient, we present the evaluation results based on the remaining features in Tables 8 and 9, with a correlation of 50 and 90 percent, respectively. Despite removing 470 and 320 features from the aggregated dataset, the proposed model demonstrates an accuracy of 100\%, proving its superior capability compared to other methods.

%table 7,8,9
\begin{table}[h]
\centering
\caption{Experimental results for final dataset.}
\label{tab:table7}
\resizebox{\textwidth}{!}{%
\begin{tabular}{|
>{\columncolor[HTML]{F4B084}}c |ccccc|}
\hline
\cellcolor[HTML]{FFFFFF}Sample size : 2.100.000 & \multicolumn{5}{c|}{\cellcolor[HTML]{8EA9DB}Experimental results for final dataset}                                                                                                                                                                                                       \\ \hline
\cellcolor[HTML]{FFFFFF}Test size   : 50\%      & \multicolumn{1}{c|}{\cellcolor[HTML]{8EA9DB}Arch 1}        & \multicolumn{1}{c|}{\cellcolor[HTML]{8EA9DB}Arch 2}        & \multicolumn{1}{c|}{\cellcolor[HTML]{8EA9DB}Arch 3}        & \multicolumn{1}{c|}{\cellcolor[HTML]{8EA9DB}Arch 4}        & \cellcolor[HTML]{8EA9DB}Arch 5        \\ \hline
\textbf{Data Standardization}                   & \multicolumn{1}{c|}{yes}                                   & \multicolumn{1}{c|}{yes}                                   & \multicolumn{1}{c|}{yes}                                   & \multicolumn{1}{c|}{yes}                                   & yes                                   \\ \hline
\textbf{Conv1D Filter size}                     & \multicolumn{1}{c|}{{[}32,16{]}}                           & \multicolumn{1}{c|}{{[}32,16{]}}                           & \multicolumn{1}{c|}{{[}32{]}}                              & \multicolumn{1}{c|}{{[}16{]}}                              & {[}8{]}                               \\ \hline
\textbf{Activation Function}                    & \multicolumn{1}{c|}{{[}'relu', 'relu'{]}}                  & \multicolumn{1}{c|}{{[}'relu', 'relu'{]}}                  & \multicolumn{1}{c|}{{[}'relu'{]}}                          & \multicolumn{1}{c|}{{[}'relu'{]}}                          & {[}'relu'{]}                          \\ \hline
\textbf{Kernel size}                            & \multicolumn{1}{c|}{{[}7,3{]}}                             & \multicolumn{1}{c|}{{[}7,3{]}}                             & \multicolumn{1}{c|}{{[}3{]}}                               & \multicolumn{1}{c|}{{[}3{]}}                               & {[}3{]}                               \\ \hline
\textbf{Max Pool 1D}                            & \multicolumn{1}{c|}{{[}2{]}}                               & \multicolumn{1}{c|}{{[}2{]}}                               & \multicolumn{1}{c|}{{[}2{]}}                               & \multicolumn{1}{c|}{{[}2{]}}                               & {[}2{]}                               \\ \hline
\textbf{Dropout}                                & \multicolumn{1}{c|}{{[}50{]}}                              & \multicolumn{1}{c|}{{[}50{]}}                              & \multicolumn{1}{c|}{{[}50{]}}                              & \multicolumn{1}{c|}{{[}50{]}}                              & {[}50{]}                              \\ \hline
\textbf{Dense}                                  & \multicolumn{1}{c|}{256-64-5}                              & \multicolumn{1}{c|}{256-64-5}                              & \multicolumn{1}{c|}{256-64-5}                              & \multicolumn{1}{c|}{256-5}                                 & 64-5                                  \\ \hline
\textbf{Activation Function}                    & \multicolumn{1}{c|}{{[}'relu', 'relu', 'softmax'{]}}       & \multicolumn{1}{c|}{{[}'relu', 'relu', 'softmax'{]}}       & \multicolumn{1}{c|}{{[}'relu', 'relu', 'softmax'{]}}       & \multicolumn{1}{c|}{{[}'relu', 'softmax'{]}}               & {[}'relu', 'softmax'{]}               \\ \hline
\textbf{Dropout}                                & \multicolumn{1}{c|}{{[}50, 50{]}}                          & \multicolumn{1}{c|}{{[}50, 50{]}}                          & \multicolumn{1}{c|}{{[}50, 50{]}}                          & \multicolumn{1}{c|}{{[}50{]}}                              & {[}50{]}                              \\ \hline
\textbf{BatchNormalization}                     & \multicolumn{1}{c|}{no}                                    & \multicolumn{1}{c|}{yes}                                   & \multicolumn{1}{c|}{yes}                                   & \multicolumn{1}{c|}{yes}                                   & yes                                   \\ \hline
\textbf{Learning Rate}                          & \multicolumn{1}{c|}{0.001}                                 & \multicolumn{1}{c|}{0.001}                                 & \multicolumn{1}{c|}{0.001}                                 & \multicolumn{1}{c|}{0.001}                                 & 0.001                                 \\ \hline
\textbf{Batch size}                             & \multicolumn{1}{c|}{500}                                   & \multicolumn{1}{c|}{500}                                   & \multicolumn{1}{c|}{500}                                   & \multicolumn{1}{c|}{500}                                   & 500                                   \\ \hline
\textbf{Epoch}                                  & \multicolumn{1}{c|}{10}                                    & \multicolumn{1}{c|}{10}                                    & \multicolumn{1}{c|}{10}                                    & \multicolumn{1}{c|}{10}                                    & 10                                    \\ \hline
\cellcolor[HTML]{8EA9DB}\textbf{accuracy}       & \multicolumn{1}{c|}{1}                                     & \multicolumn{1}{c|}{1}                                     & \multicolumn{1}{c|}{1}                                     & \multicolumn{1}{c|}{1}                                     & 0.999904                              \\ \hline
\cellcolor[HTML]{8EA9DB}\textbf{loss}           & \multicolumn{1}{c|}{2.73E-07}                              & \multicolumn{1}{c|}{3.63E-07}                              & \multicolumn{1}{c|}{5.95E-06}                              & \multicolumn{1}{c|}{2.60E-06}                              & 0.0026                                \\ \hline
\cellcolor[HTML]{8EA9DB}\textbf{Precision}      & \multicolumn{1}{c|}{1}                                     & \multicolumn{1}{c|}{1}                                     & \multicolumn{1}{c|}{1}                                     & \multicolumn{1}{c|}{1}                                     & 0.999904                              \\ \hline
\cellcolor[HTML]{8EA9DB}\textbf{Recall}         & \multicolumn{1}{c|}{1}                                     & \multicolumn{1}{c|}{1}                                     & \multicolumn{1}{c|}{1}                                     & \multicolumn{1}{c|}{1}                                     & 0.999904                              \\ \hline
\cellcolor[HTML]{8EA9DB}\textbf{F1 score}       & \multicolumn{1}{c|}{1}                                     & \multicolumn{1}{c|}{1}                                     & \multicolumn{1}{c|}{1}                                     & \multicolumn{1}{c|}{1}                                     & 0.999904                              \\ \hline
\cellcolor[HTML]{8EA9DB}\textbf{Best  Model}    & \multicolumn{1}{c|}{\cellcolor[HTML]{FFE699}HAR\_case1.h5} & \multicolumn{1}{c|}{\cellcolor[HTML]{FFE699}HAR\_case2.h5} & \multicolumn{1}{c|}{\cellcolor[HTML]{FFE699}HAR\_case3.h5} & \multicolumn{1}{c|}{\cellcolor[HTML]{FFE699}HAR\_case4.h5} & \cellcolor[HTML]{FFE699}HAR\_case5.h5 \\ \hline
\end{tabular}%
}
\end{table}

\begin{table}[h]
\centering
\caption{Experimental results for final dataset with 50\% correlation}
\label{tab:table8}
\resizebox{\textwidth}{!}{%
\begin{tabular}{|
>{\columncolor[HTML]{F4B084}}c |ccccc|}
\hline
\cellcolor[HTML]{FFFFFF}Sample size :   2.100.000 & \multicolumn{5}{c|}{\cellcolor[HTML]{8EA9DB}\textbf{Experimental results for final dataset with 0.5   correlation}}                                                                                                                                                                            \\ \hline
\cellcolor[HTML]{FFFFFF}Test size : 50\%          & \multicolumn{1}{c|}{\cellcolor[HTML]{8EA9DB}Arch 1}         & \multicolumn{1}{c|}{\cellcolor[HTML]{8EA9DB}Arch 2}         & \multicolumn{1}{c|}{\cellcolor[HTML]{8EA9DB}Arch 3}         & \multicolumn{1}{c|}{\cellcolor[HTML]{8EA9DB}Arch 4}         & \cellcolor[HTML]{8EA9DB}Arch 5         \\ \hline
\textbf{Data   Standardization}                   & \multicolumn{1}{c|}{yes}                                    & \multicolumn{1}{c|}{yes}                                    & \multicolumn{1}{c|}{yes}                                    & \multicolumn{1}{c|}{yes}                                    & yes                                    \\ \hline
\textbf{Conv1D   Filter size}                     & \multicolumn{1}{c|}{{[}32,16{]}}                            & \multicolumn{1}{c|}{{[}32,16{]}}                            & \multicolumn{1}{c|}{{[}32{]}}                               & \multicolumn{1}{c|}{{[}16{]}}                               & {[}8{]}                                \\ \hline
\textbf{Activation   Function}                    & \multicolumn{1}{c|}{{[}'relu', 'relu'{]}}                   & \multicolumn{1}{c|}{{[}'relu', 'relu'{]}}                   & \multicolumn{1}{c|}{{[}'relu'{]}}                           & \multicolumn{1}{c|}{{[}'relu'{]}}                           & {[}'relu'{]}                           \\ \hline
\textbf{Kernel   size}                            & \multicolumn{1}{c|}{{[}7,3{]}}                              & \multicolumn{1}{c|}{{[}7,3{]}}                              & \multicolumn{1}{c|}{{[}3{]}}                                & \multicolumn{1}{c|}{{[}3{]}}                                & {[}3{]}                                \\ \hline
\textbf{Max   Pool 1D}                            & \multicolumn{1}{c|}{{[}2{]}}                                & \multicolumn{1}{c|}{{[}2{]}}                                & \multicolumn{1}{c|}{{[}2{]}}                                & \multicolumn{1}{c|}{{[}2{]}}                                & {[}2{]}                                \\ \hline
\textbf{Dropout}                                  & \multicolumn{1}{c|}{{[}50{]}}                               & \multicolumn{1}{c|}{{[}50{]}}                               & \multicolumn{1}{c|}{{[}50{]}}                               & \multicolumn{1}{c|}{{[}50{]}}                               & {[}50{]}                               \\ \hline
\textbf{Dense}                                    & \multicolumn{1}{c|}{256-64-5}                               & \multicolumn{1}{c|}{256-64-5}                               & \multicolumn{1}{c|}{256-64-5}                               & \multicolumn{1}{c|}{256-5}                                  & 64-5                                   \\ \hline
\textbf{Activation   Function}                    & \multicolumn{1}{c|}{{[}'relu', 'relu', 'softmax'{]}}        & \multicolumn{1}{c|}{{[}'relu', 'relu', 'softmax'{]}}        & \multicolumn{1}{c|}{{[}'relu', 'relu', 'softmax'{]}}        & \multicolumn{1}{c|}{{[}'relu', 'softmax'{]}}                & {[}'relu', 'softmax'{]}                \\ \hline
\textbf{Dropout}                                  & \multicolumn{1}{c|}{{[}50, 50{]}}                           & \multicolumn{1}{c|}{{[}50, 50{]}}                           & \multicolumn{1}{c|}{{[}50, 50{]}}                           & \multicolumn{1}{c|}{{[}50{]}}                               & {[}50{]}                               \\ \hline
\textbf{BatchNormalization}                       & \multicolumn{1}{c|}{no}                                     & \multicolumn{1}{c|}{yes}                                    & \multicolumn{1}{c|}{yes}                                    & \multicolumn{1}{c|}{yes}                                    & yes                                    \\ \hline
\textbf{Learning   Rate}                          & \multicolumn{1}{c|}{0.001}                                  & \multicolumn{1}{c|}{0.001}                                  & \multicolumn{1}{c|}{0.001}                                  & \multicolumn{1}{c|}{0.001}                                  & 0.001                                  \\ \hline
\textbf{Batch   size}                             & \multicolumn{1}{c|}{500}                                    & \multicolumn{1}{c|}{500}                                    & \multicolumn{1}{c|}{500}                                    & \multicolumn{1}{c|}{500}                                    & 500                                    \\ \hline
\textbf{Epoch}                                    & \multicolumn{1}{c|}{10}                                     & \multicolumn{1}{c|}{10}                                     & \multicolumn{1}{c|}{10}                                     & \multicolumn{1}{c|}{10}                                     & 10                                     \\ \hline
\cellcolor[HTML]{8EA9DB}\textbf{accuracy}         & \multicolumn{1}{c|}{1}                                      & \multicolumn{1}{c|}{1}                                      & \multicolumn{1}{c|}{0.999998}                               & \multicolumn{1}{c|}{0.999997}                               & 0.992802                               \\ \hline
\cellcolor[HTML]{8EA9DB}\textbf{loss}             & \multicolumn{1}{c|}{3.14E-05}                               & \multicolumn{1}{c|}{7.92E-05}                               & \multicolumn{1}{c|}{2.54E-05}                               & \multicolumn{1}{c|}{2.43E-04}                               & 0.0342                                 \\ \hline
\cellcolor[HTML]{8EA9DB}\textbf{Precision}        & \multicolumn{1}{c|}{1}                                      & \multicolumn{1}{c|}{1}                                      & \multicolumn{1}{c|}{0.999998}                               & \multicolumn{1}{c|}{0.999997}                               & 0.992802                               \\ \hline
\cellcolor[HTML]{8EA9DB}\textbf{Recall}           & \multicolumn{1}{c|}{1}                                      & \multicolumn{1}{c|}{1}                                      & \multicolumn{1}{c|}{0.999998}                               & \multicolumn{1}{c|}{0.999997}                               & 0.992802                               \\ \hline
\cellcolor[HTML]{8EA9DB}\textbf{F1   score}       & \multicolumn{1}{c|}{1}                                      & \multicolumn{1}{c|}{1}                                      & \multicolumn{1}{c|}{0.999998}                               & \multicolumn{1}{c|}{0.999997}                               & 0.992802                               \\ \hline
\cellcolor[HTML]{8EA9DB}\textbf{Best  Model}      & \multicolumn{1}{c|}{\cellcolor[HTML]{FFE699}HAR\_corr51.h5} & \multicolumn{1}{c|}{\cellcolor[HTML]{FFE699}HAR\_corr52.h5} & \multicolumn{1}{c|}{\cellcolor[HTML]{FFE699}HAR\_corr53.h5} & \multicolumn{1}{c|}{\cellcolor[HTML]{FFE699}HAR\_corr54.h5} & \cellcolor[HTML]{FFE699}HAR\_corr55.h5 \\ \hline
\end{tabular}%
}
\end{table}

\begin{table}[h]
\centering
\caption{Experimental results for final dataset with 90\% correlation}
\label{tab:table9}
\resizebox{\textwidth}{!}{%
\begin{tabular}{|
>{\columncolor[HTML]{F4B084}}c |ccccc|}
\hline
\cellcolor[HTML]{FFFFFF}Sample size : 2.100.000 & \multicolumn{5}{c|}{\cellcolor[HTML]{8EA9DB}\textbf{Experimental results for final dataset with 0.9 correlation}}                                                                                                                                                                              \\ \hline
\cellcolor[HTML]{FFFFFF}Test size   : 50\%      & \multicolumn{1}{c|}{\cellcolor[HTML]{8EA9DB}Arch 1}         & \multicolumn{1}{c|}{\cellcolor[HTML]{8EA9DB}Arch 2}         & \multicolumn{1}{c|}{\cellcolor[HTML]{8EA9DB}Arch 3}         & \multicolumn{1}{c|}{\cellcolor[HTML]{8EA9DB}Arch 4}         & \cellcolor[HTML]{8EA9DB}Arch 5         \\ \hline
\textbf{Data Standardization}                   & \multicolumn{1}{c|}{yes}                                    & \multicolumn{1}{c|}{yes}                                    & \multicolumn{1}{c|}{yes}                                    & \multicolumn{1}{c|}{yes}                                    & yes                                    \\ \hline
\textbf{Conv1D Filter size}                     & \multicolumn{1}{c|}{{[}32,16{]}}                            & \multicolumn{1}{c|}{{[}32,16{]}}                            & \multicolumn{1}{c|}{{[}32{]}}                               & \multicolumn{1}{c|}{{[}16{]}}                               & {[}8{]}                                \\ \hline
\textbf{Activation Function}                    & \multicolumn{1}{c|}{{[}'relu', 'relu'{]}}                   & \multicolumn{1}{c|}{{[}'relu', 'relu'{]}}                   & \multicolumn{1}{c|}{{[}'relu'{]}}                           & \multicolumn{1}{c|}{{[}'relu'{]}}                           & {[}'relu'{]}                           \\ \hline
\textbf{Kernel size}                            & \multicolumn{1}{c|}{{[}7,3{]}}                              & \multicolumn{1}{c|}{{[}7,3{]}}                              & \multicolumn{1}{c|}{{[}3{]}}                                & \multicolumn{1}{c|}{{[}3{]}}                                & {[}3{]}                                \\ \hline
\textbf{Max Pool 1D}                            & \multicolumn{1}{c|}{{[}2{]}}                                & \multicolumn{1}{c|}{{[}2{]}}                                & \multicolumn{1}{c|}{{[}2{]}}                                & \multicolumn{1}{c|}{{[}2{]}}                                & {[}2{]}                                \\ \hline
\textbf{Dropout}                                & \multicolumn{1}{c|}{{[}50{]}}                               & \multicolumn{1}{c|}{{[}50{]}}                               & \multicolumn{1}{c|}{{[}50{]}}                               & \multicolumn{1}{c|}{{[}50{]}}                               & {[}50{]}                               \\ \hline
\textbf{Dense}                                  & \multicolumn{1}{c|}{256-64-5}                               & \multicolumn{1}{c|}{256-64-5}                               & \multicolumn{1}{c|}{256-64-5}                               & \multicolumn{1}{c|}{256-5}                                  & 64-5                                   \\ \hline
\textbf{Activation Function}                    & \multicolumn{1}{c|}{{[}'relu', 'relu', 'softmax'{]}}        & \multicolumn{1}{c|}{{[}'relu', 'relu', 'softmax'{]}}        & \multicolumn{1}{c|}{{[}'relu', 'relu', 'softmax'{]}}        & \multicolumn{1}{c|}{{[}'relu', 'softmax'{]}}                & {[}'relu', 'softmax'{]}                \\ \hline
\textbf{Dropout}                                & \multicolumn{1}{c|}{{[}50, 50{]}}                           & \multicolumn{1}{c|}{{[}50, 50{]}}                           & \multicolumn{1}{c|}{{[}50, 50{]}}                           & \multicolumn{1}{c|}{{[}50{]}}                               & {[}50{]}                               \\ \hline
\textbf{BatchNormalization}                     & \multicolumn{1}{c|}{no}                                     & \multicolumn{1}{c|}{yes}                                    & \multicolumn{1}{c|}{yes}                                    & \multicolumn{1}{c|}{yes}                                    & yes                                    \\ \hline
\textbf{Learning Rate}                          & \multicolumn{1}{c|}{0.001}                                  & \multicolumn{1}{c|}{0.001}                                  & \multicolumn{1}{c|}{0.001}                                  & \multicolumn{1}{c|}{0.001}                                  & 0.001                                  \\ \hline
\textbf{Batch size}                             & \multicolumn{1}{c|}{500}                                    & \multicolumn{1}{c|}{500}                                    & \multicolumn{1}{c|}{500}                                    & \multicolumn{1}{c|}{500}                                    & 500                                    \\ \hline
\textbf{Epoch}                                  & \multicolumn{1}{c|}{10}                                     & \multicolumn{1}{c|}{10}                                     & \multicolumn{1}{c|}{10}                                     & \multicolumn{1}{c|}{10}                                     & 10                                     \\ \hline
\cellcolor[HTML]{8EA9DB}\textbf{accuracy}       & \multicolumn{1}{c|}{1}                                      & \multicolumn{1}{c|}{1}                                      & \multicolumn{1}{c|}{0.999998}                               & \multicolumn{1}{c|}{1}                                      & 0.999789                               \\ \hline
\cellcolor[HTML]{8EA9DB}\textbf{loss}           & \multicolumn{1}{c|}{4.58E-06}                               & \multicolumn{1}{c|}{3.75E-06}                               & \multicolumn{1}{c|}{7.66E-06}                               & \multicolumn{1}{c|}{3.29E-05}                               & 0.0072                                 \\ \hline
\cellcolor[HTML]{8EA9DB}\textbf{Precision}      & \multicolumn{1}{c|}{1}                                      & \multicolumn{1}{c|}{1}                                      & \multicolumn{1}{c|}{0.999998}                               & \multicolumn{1}{c|}{1}                                      & 0.999789                               \\ \hline
\cellcolor[HTML]{8EA9DB}\textbf{Recall}         & \multicolumn{1}{c|}{1}                                      & \multicolumn{1}{c|}{1}                                      & \multicolumn{1}{c|}{0.999998}                               & \multicolumn{1}{c|}{1}                                      & 0.999789                               \\ \hline
\cellcolor[HTML]{8EA9DB}\textbf{F1 score}       & \multicolumn{1}{c|}{1}                                      & \multicolumn{1}{c|}{1}                                      & \multicolumn{1}{c|}{0.999998}                               & \multicolumn{1}{c|}{1}                                      & 0.999789                               \\ \hline
\cellcolor[HTML]{8EA9DB}\textbf{Best  Model}    & \multicolumn{1}{c|}{\cellcolor[HTML]{FFE699}HAR\_corr91.h5} & \multicolumn{1}{c|}{\cellcolor[HTML]{FFE699}HAR\_corr92.h5} & \multicolumn{1}{c|}{\cellcolor[HTML]{FFE699}HAR\_corr93.h5} & \multicolumn{1}{c|}{\cellcolor[HTML]{FFE699}HAR\_corr94.h5} & \cellcolor[HTML]{FFE699}HAR\_corr95.h5 \\ \hline
\end{tabular}%
}
\end{table}
\vspace{75px}
Figure 8 shows the confusion matrix for the classification of 5 activity classes: standing, crawling, walking, ascending and descending. 

\begin{figure}[h]
\centering
	\includegraphics[width=0.75\columnwidth, keepaspectratio]{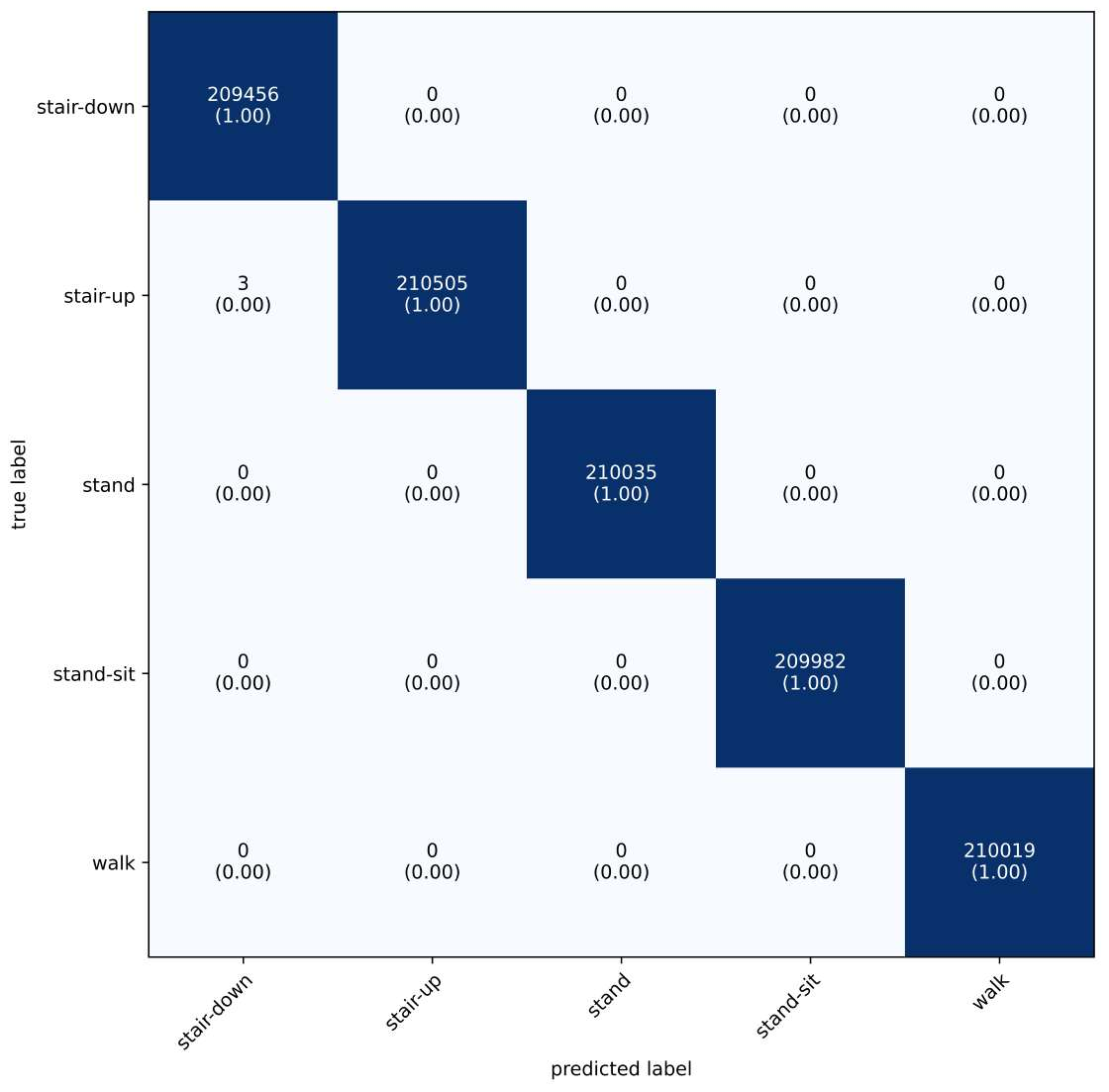}
	\captionsetup{width=1\textwidth}
	\caption{Confusion matrix for activity classes}
	\label{fig:figure8}
\end{figure}

\section*{Comparison}
Table 10 illustrates the comparison of the test results of our proposed method. We compare the results of the proposed IHARDS-CNN method with previous methods regarding F-Score and accuracy, precision, recall. The evaluation result of the proposed method demonstrates higher efficiency compared to other architectures, such as (M.R-HAR-CNN\cite{wan2020deep}, DCNN-LSTM\cite{jameer2023dcnn}, Channel-equalization-HAR\cite{huang2022channel}, and MS-IE-HHAR\cite{abdel2020deep}).

\section*{Computational Complexity}
The time complexity of a 1D Convolutional Neural Network (CNN) can be expressed in terms of the number of operations performed per layer, which depends on factors such as the number of filters, kernel size, and input size. For a single convolutional layer with (f) filters, (n) input elements, kernel size of (k), and a depth of dimension (d), the time complexity can be represented as:

\begin{eqnarray}
O(nkdf)
\end{eqnarray}

And Also, the time complexity of a neural network depends on several factors, including the architecture of the network, the number of layers, the number of neurons in each layer, and the size of the input data. For a neural network with four layers of neurons (i, j, k, l), training samples (s), and epochs (e), the time complexity can be represented as:

\begin{eqnarray}
O(es*(ij + jk + kl))
\end{eqnarray}

\section*{Conclusion}
This research proposes a new IHARDS-CNN technique and a new dataset for detecting HAR. We named the proposed dataset IHARDS, which we compiled by combining three datasets: UCI-HAR, WISDM, and KU-HAR. After conducting statistical tests on this dataset, we confirmed its accuracy. Therefore, our proposed approach has increased the speed and efficiency of the architecture due to the uniformity of the dataset, the simpler algorithm, and the absence of data aggregation in the output. Additionally, we used a simple one-dimensional CNN with several regulatory techniques in this method. Accurately and simultaneously removing the correlation features above 50 and 90 percent of the data proves its superiority over existing methods. The evaluation result of our proposed approach demonstrates accuracy, f-measure, and recall close to 100\%. Due to a slight error of two units in the confusion matrix, the accuracy becomes 100\% after rounding. 

\section*{Discussion}
Deep learning facilitated the extraction of high-level features with high accuracy, confirming the effectiveness of our proposed approach in the evaluation results. Analyzing the evaluation results of our integrated dataset compared to previous methods with separate datasets indicates high accuracy and f-measure efficiency close to 100\%. Additionally, the efficiency of the above scenario, with the help of one-dimensional CNN, has led to the simplicity and speed of our proposed model. While the dataset analysis was completed relatively quickly, generating the dataset might take time due to its nested nature. In contrast to the time spent on data integration, it offers the advantages of reduced error accumulation and an integrated algorithm. Our proposed approach is limited because it is only suitable for sensor data and cannot be used in image or video processing. Other researchers can find our proposed dataset for this research on GitHub.

\section*{Future work}
We hope future research will explore combining deep learning algorithms with other health and medical diagnosis methods. Other applications of human activity recognition in various fields, such as physiological data analysis to diagnose or prevent diseases, and dynamic environments, such as passenger transportation and fitness, could be effective in the future. Based on the potential of our proposed approach, three different fields can implement this research. Researchers can use this approach in future works in an investigative, scientific, and practical manner. They can also study the proposed unified dataset in black and white or colored image processing topics. The applications of our proposed approach can benefit various industries, such as security, health, sports, and medicine. Additionally, this approach can play a significant role in caring for older adults and young children.

%ref

\end{document}